\documentclass[aps,prd,twocolumn,superscriptaddress,nofootinbib]{revtex4-1}

\usepackage{latexsym}
\usepackage{amsmath}
\usepackage{amssymb}
\usepackage{amsfonts}
\usepackage{bm}

\usepackage{color}
\definecolor{purple}{rgb}{0.5,0,0.5}
\definecolor{blue}{rgb}{0.0,0,0.9}
\definecolor{prdblue}{rgb}{0.133,0.118,0.498}
\usepackage[colorlinks=true, pdfstartview=FitV, linkcolor=prdblue, citecolor= prdblue, urlcolor=prdblue]{hyperref}

\usepackage{supertabular} 
\usepackage{placeins}
\usepackage{epsfig}
\usepackage{graphicx}

\begin{document}


\title{The $\mathbf{\bar{q}q\bar{s}Q}$ $\mathbf{(q=u,\,d;\,Q=c,\,b)}$ tetraquark system in a chiral quark model}


\author{Gang Yang}
\email[]{yanggang@zjnu.edu.cn}
\affiliation{Department of Physics, Zhejiang Normal University, Jinhua 321004, China}

\author{Jialun Ping}
\email[]{jlping@njnu.edu.cn}
\affiliation{Department of Physics and Jiangsu Key Laboratory for Numerical Simulation of Large Scale Complex Systems, Nanjing Normal University, Nanjing 210023, P. R. China}

\author{Jorge Segovia}
\email[]{jsegovia@upo.es}
\affiliation{Departamento de Sistemas F\'isicos, Qu\'imicos y Naturales, Universidad Pablo de Olavide, E-41013 Sevilla, Spain}



\begin{abstract}
Inspired by the experimentally reported $T_{c\bar{s}}(2900)$ exotic states, the $S$-wave $\bar{q}q\bar{s}Q$ $(q=u,\,d;\,Q=c,\,b)$ tetraquarks, with spin-parity $J^P=0^+$, $1^+$ and $2^+$, in both isoscalar and isovector sectors are systematically studied in a chiral quark model. The meson-meson, diquark-antidiquark and K-type arrangements of quarks, along with all possible color wave functions, are comprehensively considered. The four-body system is solved by means of a highly efficient computational approach, the Gaussian expansion method, along with a complex-scaling formulation of the problem to disentangle bound, resonance and scattering states. This theoretical framework has already been successfully applied in various tetra- and penta-quark systems. In the complete coupled-channel case, and within the complex-range formulation, several narrow resonances of $\bar{q}q\bar{s}c$ and $\bar{q}q\bar{s}b$ systems are obtained in each allowed $I(J^P)$-channels. Particularly, the $T_{c\bar{s}}(2900)$ is well identified as a $I(J^P)=1(0^+)$ $\bar{q}q\bar{s}c$ tetraquark state with a dominant molecular structure. Meanwhile, more resonances in $\bar{q}q\bar{s}c$ and $\bar{q}q\bar{s}b$ systems are also obtained within the energy regions $2.4-3.4$ GeV and $5.7-6.7$ GeV, respectively. The predicted exotic states, which are an indication of a richer color structure when going towards multiquark systems beyond mesons and baryons, are expected to be confirmed in future high-energy particle and nuclear experiments.
\end{abstract}


\maketitle


\section{Introduction}

In 2020 two charm-strange resonances $X_0(2900)$ and $X_1(2900)$, which are presumably $ud\bar{s}\bar{c}$ tetraquark candidates, were reported by the LHCb collaboration in $B^+\rightarrow D^+D^-K^+$ decays~\cite{LHCb:2020pxc, LHCb:2020bls}. Later on, in 2023, this collaboration announced two new resonant states, $T^0_{c\bar{s}}(2900)$ and $T^{++}_{c\bar{s}}(2900)$, in a combined amplitude analysis of $B^0 \rightarrow \bar{D}^0 D^+_s \pi^-$ and $B^+ \rightarrow D^- D^+_s \pi^+$ decays~\cite{LHCb:2022sfr, LHCb:2022lzp}. These observations indicated that the signals may be open-charm tetraquark candidates with minimal quark content $c\bar{s}q\bar{q}$ $(q=u,\,d)$. Without the ability to disentangle the masses and widths of these two resonances, the LHCb collaboration determined that they are $2.908\pm0.011\pm0.020$ GeV and $0.136\pm0.023\pm0.011$ GeV in both cases; moreover, their quantum numbers $I(J^P)$ were determined to be $1(0^+)$ for both of them.

These observations trigger enormous theoretical investigations. Generally, the $T_{c\bar{s}}(2900)$ states can be well identified as a molecular structures in the $I(J^P)=1(0^+)$ channel by phenomenological models such as those presented in Refs.~\cite{Ortega:2023azl, Duan:2023lcj, Wang:2023hpp}. However, interpretations of compact configurations~\cite{Wei:2022wtr, Yang:2023evp}, threshold effects~\cite{Molina:2023ghu} and triangle anomalies~\cite{Ke:2022ocs} have also been proposed by various effective field theory approaches and quark models based on the color flux-tube. Meanwhile, strong decay properties~\cite{Yue:2022mnf, Yue:2023qgx, Lian:2023cgs} and production mechanisms~\cite{Lyu:2023jos, Huang:2023fvj, Duan:2023qsg, Lyu:2023aqn} of the exotic states have been investigated too. In addition, more $T_{c\bar{s}}$ resonances within an energy region $2.1-3.0$ GeV are predicted theoretically~\cite{Ortega:2023azl, Wei:2022wtr, Wang:2023hpp}. 

Besides the mentioned $T_{cs}$ and $T_{c\bar s}$ states dozens of exotic hadrons, whose masses are generally near the threshold of two conventional heavy flavored hadrons, have been reported experimentally within the last 20 years. At the rate at which they were being discovered, great theoretical interest and ingenious efforts with a wide variety of approaches were developed by theorists in order to reveal the nature of the unexpected exotic states, which are overall good candidates of multiquark systems. Particularly, many extensive reviews~\cite{Dong:2020hxe, Chen:2016qju, Chen:2016spr, Guo:2017jvc, Liu:2019zoy, Yang:2020atz, Dong:2021bvy, Chen:2021erj, Cao:2023rhu, Mai:2022eur, Meng:2022ozq, Chen:2022asf, Guo:2022kdi, Ortega:2020tng, Huang:2023jec, Lebed:2023vnd, Zou:2021sha, Du:2021fmf}, which explain in detail a particular theoretical method and thus capturing a certain interpretation of exotic hadrons, can be found in the literature.

Herein, we perform a comprehensive investigation of the $\bar{q}q\bar{s}Q$ $(q=u,\,d;\,Q=c,\,b)$ tetraquark systems with spin-parity $J^P=0^+$, $1^+$ and $2^+$, and in the isospin $I=0$ and $1$ sector. A variational formalism based on a highly efficient numerical approach named the Gaussian expansion method (GEM)~\cite{Hiyama:2003cu} is used to solve the 4-body Hamiltonian, which is based on a chiral quark model that has been used to describe reasonably well various tetra- and penta-quark systems~\cite{Yang:2015bmv, Yang:2018oqd, Yang:2019itm, Yang:2020fou, Yang:2020twg, Yang:2021izl, Yang:2021zhe, Yang:2022cut, Yang:2023mov, Yang:2023dzb}. Moreover, bound, resonant and scattering states can be well disentangled by solving the complex scaled Schr\"odinger equation formulated under the complex scaling method (CSM). Furthermore, the meson-meson, diquark-antidiquark and K-type arrangements of quarks, plus their couplings with all possible color wave functions, are considered. The nature of the $T_{c\bar{s}}(2900)$ states is revealed, along with the prediction of more $T_{c\bar{s}}$ and $T_{b\bar{s}}$ resonances.

We arrange this manuscript as follows. We present in Sec.~\ref{sec:model} the theoretical framework that includes a brief description of the chiral quark model and the $\bar{q}q\bar{s}Q$ tetraquark wave functions. Section~\ref{sec:results} is devoted to the analysis and discussion of the calculated results. Finally, a summary is presented in Sec.~\ref{sec:summary}.


\section{Theoretical framework}
\label{sec:model}

Phenomenological models continue to be the main tools to shed some light on the nature of multiquark candidates observed experimentally. Hence, the $\bar{q}q\bar{s}Q$ tetraquark systems are systematically investigated by means of a chiral quark model. Moreover, a high accurate computing approach for the few-body system, the Gaussian expansion method (GEM), along with a powerful complex scaling method (CSM) are adopted for investigating the bound, resonant and scattering states of the multiquark system. The theoretical formalism employed herein was described in detail in Ref.~\cite{Yang:2020atz}, and we shall then focus herein on the most relevant features of the model and the numerical approach concerning the $\bar{q}q\bar{s}Q$ tetraquarks.


\subsection{The Hamiltonian}

The four-body problem is studied in a complex scaled Schr\"odinger equation:
\begin{equation}\label{CSMSE}
\left[ H(\theta)-E(\theta) \right] \Psi_{JM}(\theta)=0 \,,
\end{equation}
where the general form of the four-body Hamiltonian for a QCD-inspired chiral quark model reads as
\begin{equation}
H(\theta) = \sum_{i=1}^{4}\left( m_i+\frac{\vec{p\,}^2_i}{2m_i}\right) - T_{\text{CM}} + \sum_{j>i=1}^{4} V(\vec{r}_{ij} e^{i\theta}) \,,
\label{eq:Hamiltonian}
\end{equation}
where $m_{i}$ is the constituent quark mass, $\vec{p}_i$ is the momentum of a quark, $T_{\text{CM}}$ is the center-of-mass kinetic energy and the last term is the two-body potential. By introducing an artificial parameter in the Hamiltonian, named the rotated angle $\theta$, three kinds of complex eigenvalues can be found, \emph{viz.} bound, resonance and scattering states can be simultaneously studied. Particularly, bound and resonance states are independent of the rotated angle $\theta$, with the first ones always placed on the real-axis of the complex energy plane, and the second ones located above the threshold line with a total decay width $\Gamma=-2\,\text{Im}(E)$. Meanwhile, the energy dots corresponding to scattering states are unstable under rotations of $\theta$ and align along the threshold line which also changes with different values of $\theta$.

The dynamics of $\bar{q}q\bar{s}Q$ tetraquark systems is driven by two-body complex scaled potentials,
\begin{equation}
\label{CQMV}
V(\vec{r}_{ij} e^{i\theta}) = V_{\chi}(\vec{r}_{ij} e^{i\theta}) + V_{\text{CON}}(\vec{r}_{ij} e^{i\theta}) + V_{\text{OGE}}(\vec{r}_{ij} e^{i\theta})  \,.
\end{equation}
In particular, dynamical chiral symmetry breaking, color-confinement and perturbative one-gluon exchange interactions are taken into account which are considered the most relevant features of QCD at its low energy regime. Since the low-lying $S$-wave positive parity $\bar{q}q\bar{s}Q$ tetraquark states shall be investigated herein, only the central and spin-spin terms of the potential are considered.

One consequence of the dynamical breaking of chiral symmetry is that Goldstone boson exchange interactions appear between constituent light quarks $u$, $d$ and $s$. Accordingly, the complex scaled chiral interaction can be written as:
\begin{equation}
V_{\chi}(\vec{r}_{ij} e^{i\theta}) = V_{\pi}(\vec{r}_{ij} e^{i\theta})+ V_{\sigma}(\vec{r}_{ij} e^{i\theta}) + V_{K}(\vec{r}_{ij} e^{i\theta}) + V_{\eta}(\vec{r}_{ij} e^{i\theta}) \,,
\end{equation}
given by
\begin{align}
&
V_{\pi}\left( \vec{r}_{ij} e^{i\theta} \right) = \frac{g_{ch}^{2}}{4\pi}
\frac{m_{\pi}^2}{12m_{i}m_{j}} \frac{\Lambda_{\pi}^{2}}{\Lambda_{\pi}^{2}-m_{\pi}
^{2}}m_{\pi} \Bigg[ Y(m_{\pi}r_{ij} e^{i\theta}) \nonumber \\
&
\hspace*{1.20cm} - \frac{\Lambda_{\pi}^{3}}{m_{\pi}^{3}}
Y(\Lambda_{\pi}r_{ij} e^{i\theta}) \bigg] (\vec{\sigma}_{i}\cdot\vec{\sigma}_{j})\sum_{a=1}^{3}(\lambda_{i}^{a}
\cdot\lambda_{j}^{a}) \,, \\
& 
V_{\sigma}\left( \vec{r}_{ij} e^{i\theta} \right) = - \frac{g_{ch}^{2}}{4\pi}
\frac{\Lambda_{\sigma}^{2}}{\Lambda_{\sigma}^{2}-m_{\sigma}^{2}}m_{\sigma} \Bigg[Y(m_{\sigma}r_{ij} e^{i\theta}) \nonumber \\
&
\hspace*{1.20cm} - \frac{\Lambda_{\sigma}}{m_{\sigma}}Y(\Lambda_{\sigma}r_{ij} e^{i\theta})
\Bigg] \,,
\end{align}
\begin{align}
& 
V_{K}\left( \vec{r}_{ij} e^{i\theta} \right)= \frac{g_{ch}^{2}}{4\pi}
\frac{m_{K}^2}{12m_{i}m_{j}}\frac{\Lambda_{K}^{2}}{\Lambda_{K}^{2}-m_{K}^{2}}m_{
K} \Bigg[ Y(m_{K}r_{ij} e^{i\theta}) \nonumber \\
&
\hspace*{1.20cm} -\frac{\Lambda_{K}^{3}}{m_{K}^{3}}Y(\Lambda_{K}r_{ij} e^{i\theta}) \Bigg] (\vec{\sigma}_{i}\cdot\vec{\sigma}_{j}) \sum_{a=4}^{7}(\lambda_{i}^{a} \cdot \lambda_{j}^{a}) \,, \\
& 
V_{\eta}\left( \vec{r}_{ij} e^{i\theta} \right) = \frac{g_{ch}^{2}}{4\pi}
\frac{m_{\eta}^2}{12m_{i}m_{j}} \frac{\Lambda_{\eta}^{2}}{\Lambda_{\eta}^{2}-m_{
\eta}^{2}}m_{\eta} \Bigg[ Y(m_{\eta}r_{ij} e^{i\theta}) \nonumber \\
&
\hspace*{1.20cm} -\frac{\Lambda_{\eta}^{3}}{m_{\eta}^{3}
}Y(\Lambda_{\eta}r_{ij} e^{i\theta}) \Bigg] (\vec{\sigma}_{i}\cdot\vec{\sigma}_{j})
\Big[\cos\theta_{p} \left(\lambda_{i}^{8}\cdot\lambda_{j}^{8}
\right) \nonumber \\
&
\hspace*{1.20cm} -\sin\theta_{p} \Big] \,,
\end{align}
where $Y(x)=e^{-x}/x$ is the Yukawa function. The physical $\eta$ meson, instead of the octet one, is considered by introducing a model parameter of angle $\theta_p$. The $\lambda^{a}$ are the SU(3) flavor Gell-Mann matrices. Taken from their experimental values, $m_{\pi}$, $m_{K}$ and $m_{\eta}$ are the masses of the SU(3) Goldstone bosons. The value of $m_{\sigma}$ is determined through the relation $m_{\sigma}^{2}\simeq m_{\pi}^{2}+4m_{u,d}^{2}$~\cite{Scadron:1982eg}. Finally, the chiral coupling constant, $g_{ch}$, is determined from the $\pi NN$ coupling constant through
\begin{equation}
\frac{g_{ch}^{2}}{4\pi}=\frac{9}{25}\frac{g_{\pi NN}^{2}}{4\pi} \frac{m_{u,d}^{2}}{m_{N}^2} \,,
\end{equation}
which assumes that flavor SU(3) is an exact symmetry only broken by the different mass of the strange quark.

Color confinement should be encoded in the non-Abelian character of QCD. On one hand, lattice-regularized QCD has demonstrated that multi-gluon exchanges produce an attractive linearly rising potential proportional to the distance between infinite-heavy quarks~\cite{Bali:2005fu}. On the other hand, the spontaneous creation of light-quark pairs from the QCD vacuum may give rise at the same scale to a breakup of the created color flux-tube~\cite{Bali:2005fu}. We can phenomenologically describe the above two observations by
\begin{equation}
V_{\text{CON}}(\vec{r}_{ij} e^{i\theta})=\left[-a_{c}(1-e^{-\mu_{c}r_{ij} e^{i\theta}})+\Delta \right] 
(\lambda_{i}^{c}\cdot \lambda_{j}^{c}) \,,
\label{eq:conf}
\end{equation}
where $\lambda^c$ denote as the SU(3) color Gell-Mann matrices, and $a_{c}$, $\mu_{c}$ and $\Delta$ are model parameters. When the rotated angle $\theta$ is $0^\circ$, one can see in Eq.~\eqref{eq:conf} that the real-range potential is linear at short inter-quark distances with an effective confinement strength $\sigma = -a_{c} \, \mu_{c} \, (\lambda^{c}_{i}\cdot \lambda^{c}_{j})$, while it becomes a constant at large distances, $V_{\text{thr.}} = (\Delta-a_c)(\lambda^{c}_{i}\cdot \lambda^{c}_{j})$.

Beyond the chiral symmetry breaking energy scale, one also expects the dynamics to be governed by perturbative effects of QCD. In particular, the one-gluon exchange potential, which includes the so-called Coulomb and color-magnetic interactions, is the leading order contribution:
\begin{align}
&
V_{\text{OGE}}(\vec{r}_{ij} e^{i\theta}) = \frac{1}{4} \alpha_{s} (\lambda_{i}^{c}\cdot \lambda_{j}^{c}) \Bigg[\frac{1}{r_{ij} e^{i\theta}} \nonumber \\ 
&
\hspace*{1.60cm} - \frac{1}{6m_{i}m_{j}} (\vec{\sigma}_{i}\cdot\vec{\sigma}_{j}) 
\frac{e^{-r_{ij} e^{i\theta} /r_{0}(\mu_{ij})}}{r_{ij} e^{i\theta} r_{0}^{2}(\mu_{ij})} \Bigg] \,,
\end{align}
where $\vec{\sigma}$ denote as the Pauli matrices, and $r_{0}(\mu_{ij})=\hat{r}_{0}/\mu_{ij}$ depends on the reduced mass of a $q\bar{q}$ pair. Besides, the regularized contact term is
\begin{equation}
\delta(\vec{r}_{ij} e^{i\theta}) \sim \frac{1}{4\pi r_{0}^{2}(\mu_{ij})}\frac{e^{-r_{ij} e^{i\theta} / r_{0}(\mu_{ij})}}{r_{ij} e^{i\theta} } \,.
\end{equation}

An effective scale-dependent strong coupling constant, $\alpha_s(\mu_{ij})$, provides a consistent description of mesons and baryons from light to heavy quark sectors. The frozen coupling constant is used of, for instance, Ref.~\cite{Segovia:2013wma},
\begin{equation}
\alpha_{s}(\mu_{ij})=\frac{\alpha_{0}}{\ln\left(\frac{\mu_{ij}^{2}+\mu_{0}^{2}}{\Lambda_{0}^{2}} \right)} \,,
\end{equation}
where $\alpha_{0}$, $\mu_{0}$ and $\Lambda_{0}$ are model parameters.

All of the discussed model parameters are summarized in Table~\ref{tab:model}. They have been fixed along the last two decades by thorough studies of hadron phenomenology such as meson~\cite{Segovia:2008zza, Segovia:2015dia, Ortega:2020uvc} and baryon~\cite{Valcarce:1995dm, Yang:2017qan, Yang:2019lsg} spectra, hadron decays and reactions~\cite{Segovia:2009zz, Segovia:2011zza, Segovia:2011dg}, coupling between conventional hadrons and hadron-hadron thresholds~\cite{Ortega:2009hj, Ortega:2016pgg, Ortega:2016hde} as well as molecular hadron-hadron formation~\cite{Ortega:2018cnm, Ortega:2021xst, Ortega:2022efc}. Furthermore, for later concern, Table~\ref{MesonMass} lists theoretical and experimental (if available) masses of $1S$ and $2S$ states of $q\bar{q}$ and $\bar{q}Q$ $(q=u,\,d,\,s;\, Q=c,\,b)$ mesons.

\begin{table}[!t]
\caption{\label{tab:model} Chiral quark model parameters.}
\begin{ruledtabular}
\begin{tabular}{llr}
Quark masses     & $m_q\,(q=u,\,d)$ (MeV) & 313 \\
                 & $m_s$ (MeV) &  555 \\
                 & $m_c$ (MeV) & 1752 \\
                 & $m_b$ (MeV) & 5100 \\[2ex]
Goldstone bosons & $\Lambda_\pi=\Lambda_\sigma~$ (fm$^{-1}$) &   4.20 \\
                 & $\Lambda_\eta=\Lambda_K$ (fm$^{-1}$)      &   5.20 \\
                 & $g^2_{ch}/(4\pi)$                         &   0.54 \\
                 & $\theta_P(^\circ)$                        & -15 \\[2ex]
Confinement      & $a_c$ (MeV)         & 430 \\
                 & $\mu_c$ (fm$^{-1})$ & 0.70 \\
                 & $\Delta$ (MeV)      & 181.10 \\[2ex]
OGE              & $\alpha_0$              & 2.118 \\
                 & $\Lambda_0~$(fm$^{-1}$) & 0.113 \\
                 & $\mu_0~$(MeV)           & 36.976 \\
                 & $\hat{r}_0~$(MeV~fm)    & 28.17 \\
\end{tabular}
\end{ruledtabular}
\end{table}

\begin{table*}[!t]
\caption{\label{MesonMass} Theoretical and experimental (if available) masses of $1S$ and $2S$ states of $q\bar{q}$ and $\bar{q}Q\,(q=u, d, s;\, Q=c,\,b)$ mesons, unit in MeV.}
\begin{ruledtabular}
\begin{tabular}{lccclccc}
Meson & $nL$ & $M_{\text{The.}}$ & $M_{\text{Exp.}}$  & Meson & $nL$ & $M_{\text{The.}}$  & $M_{\text{Exp.}}$ \\
\hline
$\pi$ & $1S$ &  $149$ & $140$  & $\eta$ & $1S$ &  $689$ & $548$ \\
    & $2S$ & $1291$ & $1300$    &  & $2S$ & $1443$ & $1295$ \\[2ex]
$\rho$ & $1S$ &  $772$ & $770$  & $\omega$ & $1S$ &  $696$ & $782$ \\
    & $2S$ & $1479$ & $1450$    &  & $2S$ & $1449$ & $1420$ \\[2ex]
$K$ & $1S$ &  $481$ & $494$  & $K^*$ & $1S$ &  $907$ & $892$ \\
    & $2S$ & $1468$ & $1460$    &  & $2S$ & $1621$ & $1630$ \\[2ex]
$D$ & $1S$ &  $1897$ & $1870$  & $D^*$ & $1S$ &  $2017$ & $2007$ \\
    & $2S$ & $2648$ & -    &  & $2S$ & $2704$ & - \\[2ex]
$D_s$ & $1S$ &  $1989$ & $1968$  & $D^*_s$ & $1S$ &  $2115$ & $2112$ \\
    & $2S$ & $2705$ & -    &  & $2S$ & $2769$ & - \\[2ex]
$B$ & $1S$ &  $5278$ & $5280$  & $B^*$ & $1S$ &  $5319$ & $5325$ \\
    & $2S$ & $5984$ & -    &  & $2S$ & $6005$ & - \\[2ex]
$B_s$ & $1S$ &  $5355$ & $5367$  & $B^*_s$ & $1S$ &  $5400$ & $5415$ \\
    & $2S$ & $6017$ & -    &  & $2S$ & $6042$ & - 
\end{tabular}
\end{ruledtabular}
\end{table*}

\begin{figure}[ht]
\epsfxsize=3.4in \epsfbox{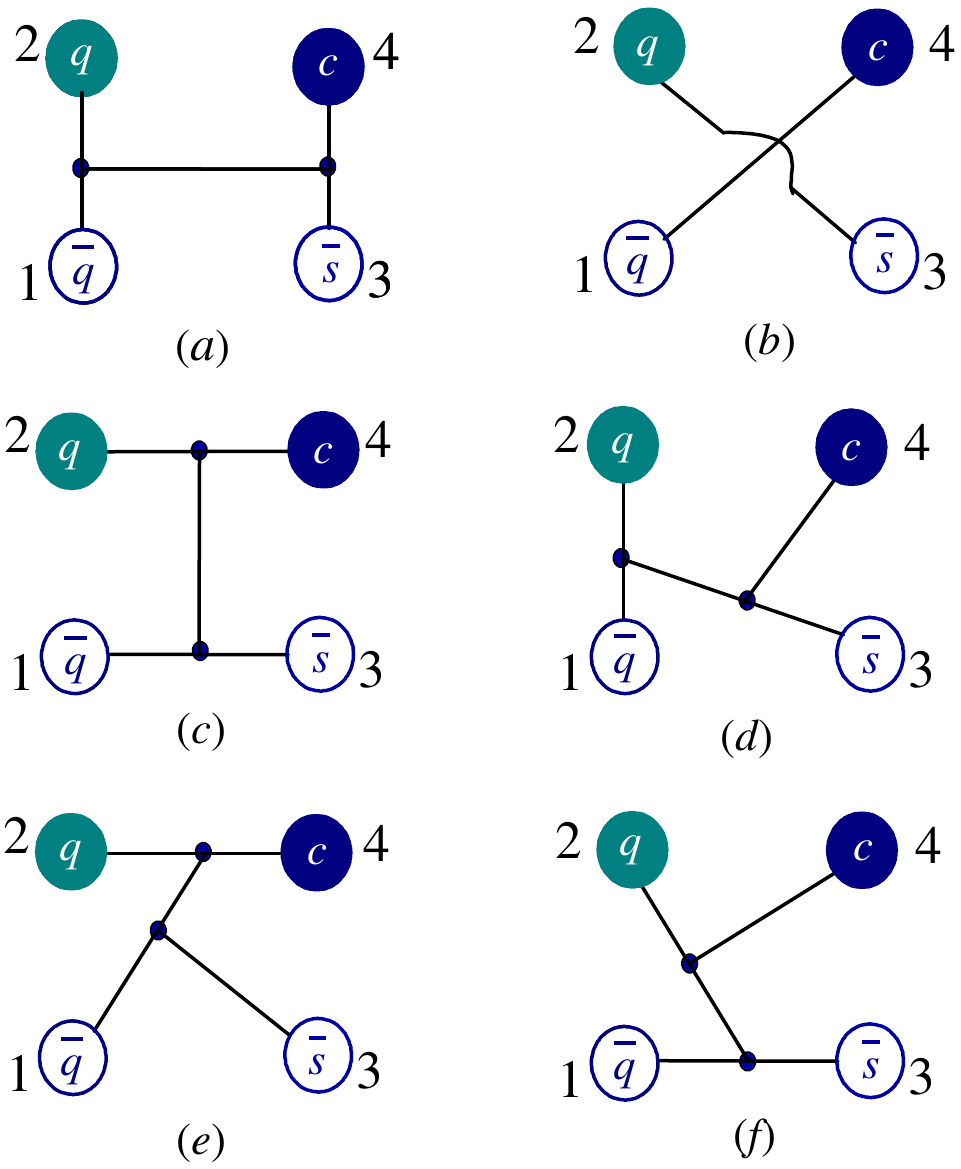}
\caption{\label{QQqq} Six types of configurations are considered for the $\bar{q}q\bar{s}Q$ $(q=u,\,d;\,Q=c,\,b)$ tetraquarks. Panels $(a)$ and $(b)$ are meson-meson structures, panel $(c)$ is diquark-antidiquark arrangement and the K-type configurations are from panel $(d)$ to $(f)$.}
\end{figure}


\subsection{The wave function}

The $S$-wave $\bar{q}q\bar{s}Q$ $(q=u,\,d;\,Q=c,\,b)$ tetraquark configurations are shown in Figure~\ref{QQqq}. Particularly, Figs.~\ref{QQqq}(a) and (b) are the meson-meson structures, Fig.~\ref{QQqq}(c) is the diquark-antidiquark arrangement, and the K-type configurations are from panels (d) to (f). For the purpose of solving a manageable $4$-body problem in the case of fully-coupled, the K-type configurations are sometimes restricted as in our previous investigations~\cite{Yang:2021zhe, Yang:2022cut}. Furthermore, it deserves to be stressed that just one configuration would be enough for the calculation, if all radial and orbital excited states were taken into account; however, this is obviously inefficient and thus a more economic way to proceed is the combination of the different mentioned structures.

At the quark level, the total wave function of a tetraquark system is the internal product of color, spin, flavor and space wave functions. Firstly, concerning the color degree-of-freedom, the colorless wave function of a $4$-quark system in meson-meson configuration can be obtained by either two coupled color-singlet clusters, $1\otimes 1$:
\begin{align}
\label{Color1}
\chi^c_1 &= \frac{1}{3}(\bar{r}r+\bar{g}g+\bar{b}b)\times (\bar{r}r+\bar{g}g+\bar{b}b) \,,
\end{align}
or two coupled color-octet clusters, $8\otimes 8$:
\begin{align}
\label{Color2}
\chi^c_2 &= \frac{\sqrt{2}}{12}(3\bar{b}r\bar{r}b+3\bar{g}r\bar{r}g+3\bar{b}g\bar{g}b+3\bar{g}b\bar{b}g+3\bar{r}g\bar{g}r
\nonumber\\
&+3\bar{r}b\bar{b}r+2\bar{r}r\bar{r}r+2\bar{g}g\bar{g}g+2\bar{b}b\bar{b}b-\bar{r}r\bar{g}g
\nonumber\\
&-\bar{g}g\bar{r}r-\bar{b}b\bar{g}g-\bar{b}b\bar{r}r-\bar{g}g\bar{b}b-\bar{r}r\bar{b}b) \,.
\end{align}
The first color state is the so-called color-singlet channel and the second one is the named hidden-color case.

The color wave functions associated to the diquark-antidiquark structure are the coupled color triplet-antitriplet clusters, $3\otimes \bar{3}$:
\begin{align}
\label{Color3}
\chi^c_3 &= \frac{\sqrt{3}}{6}(\bar{r}r\bar{g}g-\bar{g}r\bar{r}g+\bar{g}g\bar{r}r-\bar{r}g\bar{g}r+\bar{r}r\bar{b}b
\nonumber\\
&-\bar{b}r\bar{r}b+\bar{b}b\bar{r}r-\bar{r}b\bar{b}r+\bar{g}g\bar{b}b-\bar{b}g\bar{g}b
\nonumber\\
&+\bar{b}b\bar{g}g-\bar{g}b\bar{b}g) \,,
\end{align}
and the coupled color sextet-antisextet clusters, $6\otimes \bar{6}$:
\begin{align}
\label{Color4}
\chi^c_4 &= \frac{\sqrt{6}}{12}(2\bar{r}r\bar{r}r+2\bar{g}g\bar{g}g+2\bar{b}b\bar{b}b+\bar{r}r\bar{g}g+\bar{g}r\bar{r}g
\nonumber\\
&+\bar{g}g\bar{r}r+\bar{r}g\bar{g}r+\bar{r}r\bar{b}b+\bar{b}r\bar{r}b+\bar{b}b\bar{r}r
\nonumber\\
&+\bar{r}b\bar{b}r+\bar{g}g\bar{b}b+\bar{b}g\bar{g}b+\bar{b}b\bar{g}g+\bar{g}b\bar{b}g) \,.
\end{align}

Meanwhile, the possible color-singlet wave functions of three K-type structures are given by
\begin{align}
\label{Color5}
\chi^c_5 &= \frac{1}{6\sqrt{2}}(\bar{r}r\bar{r}r+\bar{g}g\bar{g}g-2\bar{b}b\bar{b}b)+
\nonumber\\
&\frac{1}{2\sqrt{2}}(\bar{r}b\bar{b}r+\bar{r}g\bar{g}r+\bar{g}b\bar{b}g+\bar{g}r\bar{r}g+\bar{b}g\bar{g}b+\bar{b}r\bar{r}b)-
\nonumber\\
&\frac{1}{3\sqrt{2}}(\bar{g}g\bar{r}r+\bar{r}r\bar{g}g)+\frac{1}{6\sqrt{2}}(\bar{b}b\bar{r}r+\bar{b}b\bar{g}g+\bar{r}r\bar{b}b+\bar{g}g\bar{b}b) \,,
\end{align}
\begin{align}
\label{Color6}
\chi^c_6 &= \chi^c_1 \,,
\end{align}
\begin{align}
\label{Color9}
\chi^c_7 &= \frac{1}{2\sqrt{6}}(\bar{r}b\bar{b}r+\bar{r}r\bar{b}b+\bar{g}b\bar{b}g+\bar{g}g\bar{b}b+\bar{r}g\bar{g}r+\bar{r}r\bar{g}g+
\nonumber\\
&\bar{b}b\bar{g}g+\bar{b}g\bar{g}b+\bar{g}g\bar{r}r+\bar{g}r\bar{r}g+\bar{b}b\bar{r}r+\bar{b}r\bar{r}b)+
\nonumber\\
&\frac{1}{\sqrt{6}}(\bar{r}r\bar{r}r+\bar{g}g\bar{g}g+\bar{b}b\bar{b}b) \,,
\end{align}
\begin{align}
\label{Color10}
\chi^c_8 &= \frac{1}{2\sqrt{3}}(\bar{r}b\bar{b}r-\bar{r}r\bar{b}b+\bar{g}b\bar{b}g-\bar{g}g\bar{b}b+\bar{r}g\bar{g}r-\bar{r}r\bar{g}g-
\nonumber\\
&\bar{b}b\bar{g}g+\bar{b}g\bar{g}b-\bar{g}g\bar{r}r+\bar{g}r\bar{r}g-\bar{b}b\bar{r}r+\bar{b}r\bar{r}b) \,,
\end{align}
\begin{align}
\label{Color11}
\chi^c_9 &= \chi^c_7 \,,
\end{align}
\begin{align}
\label{Color12}
\chi^c_{10} &= -\chi^c_8 \,.
\end{align}

As for the flavor degree-of-freedom, both iso-scalar ($I=0$) and iso-vector ($I=1$) channels of $\bar{q}q\bar{s}Q$ $(q=u,\,d;\,Q=c,\,b)$ tetraquarks should be considered. In particular, for meson-meson and part of K-type (Fig.~\ref{QQqq}(d) and (e)) configurations, the flavor wave functions, which are denoted as $\chi^{f_1}_{I, M_I}$, are
\begin{align}
&
\chi_{0,0}^{f_1} = -\frac{1}{\sqrt{2}}(\bar{u}u\bar{s}Q+\bar{d}d\bar{s}Q) \,, \\
&
\chi_{1,0}^{f_1} = \frac{1}{\sqrt{2}}(-\bar{u}u\bar{s}Q+\bar{d}d\bar{s}Q) \,.
\end{align}
While, by similar notations $\chi^{f_2}_{I, M_I}$ and $\chi^{f_3}_{I, M_I}$, where superscripts $2$ and $3$ refer to symmetry and antisymmetry properties between the $\bar{q}\bar{s}$-pair, respectively, the wave functions of diquark-antidiquark and K-type (Fig.~\ref{QQqq}(f)) structures read
\begin{align}
&
\chi_{0,0}^{f_2} = -\frac{1}{2}(\bar{u}u\bar{s}Q+\bar{s}u\bar{u}Q+\bar{d}d\bar{s}Q+\bar{s}d\bar{d}Q) \,, \\
&
\chi_{0,0}^{f_3} = +\frac{1}{2}(-\bar{u}u\bar{s}Q+\bar{s}u\bar{u}Q-\bar{d}d\bar{s}Q+\bar{s}d\bar{d}Q) \,, \\
&
\chi_{1,0}^{f_2} = +\frac{1}{2}(-\bar{u}u\bar{s}Q-\bar{s}u\bar{u}Q+\bar{d}d\bar{s}Q+\bar{s}d\bar{d}Q) \,, \\
&
\chi_{1,0}^{f_3} = +\frac{1}{2}(-\bar{u}u\bar{s}Q+\bar{s}u\bar{u}Q+\bar{d}d\bar{s}Q-\bar{s}d\bar{d}Q) \,.
\end{align}
Herein, the third component of the isospin, $M_I$, is fixed to be zero for simplicity, and this is due to the fact that there is no flavor-dependent interaction in the Hamiltonian which can distinguishes the third component of the isospin $I$.

Now let us consider the $S$-wave ground states with spin ($S$) ranging from $0$ to $2$. Therefore, the spin wave functions, $\chi^{\sigma_i}_{S, M_S}$, are given by ($M_S$ can be set to be equal to $S$ without lossing generality):
\begin{align}
\label{SWF0}
\chi_{0,0}^{\sigma_{u_1}}(4) &= \chi^\sigma_{00}\chi^\sigma_{00} \,, \\
\chi_{0,0}^{\sigma_{u_2}}(4) &= \frac{1}{\sqrt{3}}(\chi^\sigma_{11}\chi^\sigma_{1,-1}-\chi^\sigma_{10}\chi^\sigma_{10}+\chi^\sigma_{1,-1}\chi^\sigma_{11}) \,, \\
\chi_{0,0}^{\sigma_{u_3}}(4) &= \frac{1}{\sqrt{2}}\big((\sqrt{\frac{2}{3}}\chi^\sigma_{11}\chi^\sigma_{\frac{1}{2}, -\frac{1}{2}}-\sqrt{\frac{1}{3}}\chi^\sigma_{10}\chi^\sigma_{\frac{1}{2}, \frac{1}{2}})\chi^\sigma_{\frac{1}{2}, -\frac{1}{2}} \nonumber \\ 
&-(\sqrt{\frac{1}{3}}\chi^\sigma_{10}\chi^\sigma_{\frac{1}{2}, -\frac{1}{2}}-\sqrt{\frac{2}{3}}\chi^\sigma_{1, -1}\chi^\sigma_{\frac{1}{2}, \frac{1}{2}})\chi^\sigma_{\frac{1}{2}, \frac{1}{2}}\big) \,, \\
\chi_{0,0}^{\sigma_{u_4}}(4) &= \frac{1}{\sqrt{2}}(\chi^\sigma_{00}\chi^\sigma_{\frac{1}{2}, \frac{1}{2}}\chi^\sigma_{\frac{1}{2}, -\frac{1}{2}}-\chi^\sigma_{00}\chi^\sigma_{\frac{1}{2}, -\frac{1}{2}}\chi^\sigma_{\frac{1}{2}, \frac{1}{2}}) \,,
\end{align}
for $(S,M_S)=(0,0)$, by 
\begin{align}
\label{SWF1}
\chi_{1,1}^{\sigma_{w_1}}(4) &= \chi^\sigma_{00}\chi^\sigma_{11} \,, \\ 
\chi_{1,1}^{\sigma_{w_2}}(4) &= \chi^\sigma_{11}\chi^\sigma_{00} \,, \\
\chi_{1,1}^{\sigma_{w_3}}(4) &= \frac{1}{\sqrt{2}} (\chi^\sigma_{11} \chi^\sigma_{10}-\chi^\sigma_{10} \chi^\sigma_{11}) \,, \\
\chi_{1,1}^{\sigma_{w_4}}(4) &= \sqrt{\frac{3}{4}}\chi^\sigma_{11}\chi^\sigma_{\frac{1}{2}, \frac{1}{2}}\chi^\sigma_{\frac{1}{2}, -\frac{1}{2}}-\sqrt{\frac{1}{12}}\chi^\sigma_{11}\chi^\sigma_{\frac{1}{2}, -\frac{1}{2}}\chi^\sigma_{\frac{1}{2}, \frac{1}{2}} \nonumber \\ 
&-\sqrt{\frac{1}{6}}\chi^\sigma_{10}\chi^\sigma_{\frac{1}{2}, \frac{1}{2}}\chi^\sigma_{\frac{1}{2}, \frac{1}{2}} \,, \\
\chi_{1,1}^{\sigma_{w_5}}(4) &= (\sqrt{\frac{2}{3}}\chi^\sigma_{11}\chi^\sigma_{\frac{1}{2}, -\frac{1}{2}}-\sqrt{\frac{1}{3}}\chi^\sigma_{10}\chi^\sigma_{\frac{1}{2}, \frac{1}{2}})\chi^\sigma_{\frac{1}{2}, \frac{1}{2}} \,, \\
\chi_{1,1}^{\sigma_{w_6}}(4) &= \chi^\sigma_{00}\chi^\sigma_{\frac{1}{2}, \frac{1}{2}}\chi^\sigma_{\frac{1}{2}, \frac{1}{2}} \,,
\end{align}
for $(S,M_S)=(1,1)$, and by 
\begin{align}
\label{SWF2}
\chi_{2,2}^{\sigma_{1}}(4) &= \chi^\sigma_{11}\chi^\sigma_{11} \,,
\end{align}
for $(S,M_S)=(2,2)$. Particularly, superscripts $u_1,\ldots,u_4$ and $w_1,\ldots,w_6$ determine the spin wave function for each configuration of the $\bar{q}q\bar{s}Q$ tetraquark system, their values are listed in Table~\ref{SpinIndex}. Furthermore, the expressions above are obtained by considering the coupling between two sub-clusters whose spin wave functions are given by trivial SU(2) algebra and the necessary basis reads as
\begin{align}
\label{Spin}
&\chi^\sigma_{00} = \frac{1}{\sqrt{2}}(\chi^\sigma_{\frac{1}{2}, \frac{1}{2}} \chi^\sigma_{\frac{1}{2}, -\frac{1}{2}}-\chi^\sigma_{\frac{1}{2}, -\frac{1}{2}} \chi^\sigma_{\frac{1}{2}, \frac{1}{2}}) \,, \\
&\chi^\sigma_{11} = \chi^\sigma_{\frac{1}{2}, \frac{1}{2}} \chi^\sigma_{\frac{1}{2}, \frac{1}{2}} \,, \\
&\chi^\sigma_{1,-1} = \chi^\sigma_{\frac{1}{2}, -\frac{1}{2}} \chi^\sigma_{\frac{1}{2}, -\frac{1}{2}} \,, \\
&\chi^\sigma_{10} = \frac{1}{\sqrt{2}}(\chi^\sigma_{\frac{1}{2}, \frac{1}{2}} \chi^\sigma_{\frac{1}{2}, -\frac{1}{2}}+\chi^\sigma_{\frac{1}{2}, -\frac{1}{2}} \chi^\sigma_{\frac{1}{2}, \frac{1}{2}}) \,.
\end{align}

\begin{table}[!t]
\caption{\label{SpinIndex} Values of the superscripts $u_1,\ldots,u_4$ and $w_1,\ldots,w_6$ that specify the spin wave function for each configuration of the $\bar{q}q\bar{s}Q$ $(q=u,\,d;\, Q=c,\,b)$ tetraquark systems.}
\begin{ruledtabular}
\begin{tabular}{lccccc}
& Di-meson & Diquark-antidiquark & $K_1$ & $K_2$ & $K_3$ \\
\hline
$u_1$ & 1 & 3 & & &  \\
$u_2$ & 2 & 4 & & &  \\
$u_3$ &   &   & 5 & 7 &  9 \\
$u_4$ &   &   & 6 & 8 & 10  \\[2ex]
$w_1$ & 1 & 4 & & &  \\
$w_2$ & 2 & 5 & & &  \\
$w_3$ & 3 & 6 & & &  \\
$w_4$ &   &   & 7 & 10 & 13 \\
$w_5$ &   &   & 8 & 11 & 14  \\
$w_6$ &   &   & 9 & 12 & 15 
\end{tabular}
\end{ruledtabular}
\end{table}

The Rayleigh-Ritz variational principle, which is one of the most extended tools to solve eigenvalue problems, is employed to solve the Schr\"odinger-like 4-body system equation. Generally, within a complex-scaling theoretical framework, the spatial wave function is written as follows
\begin{equation}
\label{eq:WFexp}
\psi_{LM_L}= \left[ \left[ \phi_{n_1l_1}(\vec{\rho}e^{i\theta}\,) \phi_{n_2l_2}(\vec{\lambda}e^{i\theta}\,)\right]_{l} \phi_{n_3l_3}(\vec{R}e^{i\theta}\,) \right]_{L M_L} \,,
\end{equation}
where the internal Jacobi coordinates are defined as
\begin{align}
\vec{\rho} &= \vec{x}_1-\vec{x}_{2(4)} \,, \\
\vec{\lambda} &= \vec{x}_3 - \vec{x}_{4(2)} \,, \\
\vec{R} &= \frac{m_1 \vec{x}_1 + m_{2(4)} \vec{x}_{2(4)}}{m_1+m_{2(4)}}- \frac{m_3 \vec{x}_3 + m_{4(2)} \vec{x}_{4(2)}}{m_3+m_{4(2)}} \,,
\end{align}
for the meson-meson configurations of Figs.~\ref{QQqq}$(a)$ and $(b)$; and as
\begin{align}
\vec{\rho} &= \vec{x}_1-\vec{x}_3 \,, \\
\vec{\lambda} &= \vec{x}_2 - \vec{x}_4 \,, \\
\vec{R} &= \frac{m_1 \vec{x}_1 + m_3 \vec{x}_3}{m_1+m_3}- \frac{m_2 \vec{x}_2 + m_4 \vec{x}_4}{m_2+m_4} \,,
\end{align}
for the diquark-antidiquark structure of Fig.~\ref{QQqq}$(c)$. The remaining K-type configurations shown in Fig.~\ref{QQqq}$(d)$ to $(f)$ are ($i, j, k, l$ take values according to the panels $(d)$ to $(f)$ of Fig.~\ref{QQqq}):
\begin{align}
\vec{\rho} &= \vec{x}_i-\vec{x}_j \,, \\
\vec{\lambda} &= \vec{x}_k- \frac{m_i \vec{x}_i + m_j \vec{x}_j}{m_i+m_j} \,, \\
\vec{R} &= \vec{x}_l- \frac{m_i \vec{x}_i + m_j \vec{x}_j+m_k \vec{x}_k}{m_i+m_j+m_k} \,.
\end{align}
It is obvious now that the center-of-mass kinetic term $T_\text{CM}$ can be completely eliminated for a non-relativistic system defined in any of the above sets of relative motion coordinates.

The basis expansion of the genuine wave function of Eq.~(\ref{eq:WFexp}) is a crucial aspect in the Rayleigh-Ritz variational method. By employing the Gaussian expansion method (GEM)~\cite{Hiyama:2003cu}, which has proven to be quite efficient on solving the bound-state problem of multi-body systems, the spatial wave functions corresponding to the four relative motions are all expanded with Gaussian basis functions, whose sizes are taken in geometric progression. Hence, the form of orbital wave functions, $\phi$, in Eq.~\eqref{eq:WFexp} for a $S$-wave tetraquark system is simply written as 
\begin{align}
&
\phi_{nlm}(\vec{r}e^{i\theta}\,) = \sqrt{1/4\pi} \, N_{nl} \, (re^{i\theta})^{l} \, e^{-\nu_{n} (re^{i\theta})^2} \,.
\end{align}

Finally, the complete wave function, which fulfills the Pauli principle, is written as
\begin{align}
\label{TPs}
 \Psi_{J M_J, I} &= \sum_{i, j, k} c_{ijk} \Psi_{J M_J, I, i, j, k} \nonumber \\
 &=\sum_{i, j, k} c_{ijk} {\cal A} \left[ \left[ \psi_{L M_L} \chi^{\sigma_i}_{S M_S}(4) \right]_{J M_J} \chi^{f_j}_I \chi^{c}_k \right] \,,
\end{align}
where $\cal{A}$ is the anti-symmetry operator of $\bar{q}q\bar{s}Q$ tetraquark systems, which takes into account the use of SU(3) flavor symmetry. Its definition, according to Fig.~\ref{QQqq}, is 
\begin{equation}
\label{Antisym}
{\cal{A}} = 1-(13) \,.
\end{equation}
This is necessary in our theoretical framework, since the complete wave function of the 4-quark system is constructed from two sub-clusters: meson-meson, diquark-antidiquark and K-type configurations.
Furthermore, the so-called expansion coefficients, $c_{ijk}$, fulfill
\begin{align}
\label{TPs}
\vert c_{ijk} \vert^2&=\langle \Psi_{J M_J, I, i, j, k} \vert \Psi_{JM_J,I} \rangle \,,  \\
\sum_{i,j,k} \vert c_{ijk} \vert^2&=1 \,.
\end{align}
They are determined, together with the eigenenergy, by a generalized matrix eigenvalue problem.

In the next section, where computed results on the $\bar{q}q\bar{s}Q$ tetraquarks are discussed, we firstly study the systems by a real-range analysis, \emph{viz.}, the rotated angle $\theta$ is equal to $0^{\circ}$. In this case, when a complete coupled-channel calculation of matrix diagonalization is performed, possible resonant states are embedded in the continuum. However, one can employ the CSM, with appropriate non-zero values of $\theta$, to disentangle bound, resonance and scattering states in a complex energy plane. Accordingly, with the purpose of solving manageable eigevalue problems, the artificial parameter of rotated angle is ranged form $0^\circ$ to $6^\circ$. Meanwhile, with the cooperation of real- and complex-range computations, available exotic states, which are firstly obtained within a complex-range analysis, and then can be identified among continuum states according to its mass in a real-range calculation, are further investigated by analyzing their dominant quark arrangements, sizes and decay patterns.


\section{Results}
\label{sec:results}

The $S$-wave $\bar{q}q\bar{s}Q$ $(q=u,\,d;\,Q=c,\,b)$ tetraquarks are systematically studied by including meson-meson, diquark-antidiquark and K-type configurations. Therefore, the total angular momentum, $J$, coincides with the total spin, $S$, and can take values of $0$, $1$ and $2$. The parity of tetraquark system is then positive. Furthermore, both the iso-scalar $(I=0)$ and -vector $(I=1)$ sectors of $\bar{q}q\bar{s}Q$ tetraquarks are considered.

Tables~\ref{GresultCC1} to~\ref{GresultR12} list calculated results of low-lying $\bar{q}q\bar{s}Q$ tetraquark states. In particular, real-range computations on the lowest-lying masses of each tetraquark system in the allowed $I(J^P)$ quantum numbers are presented in Tables~\ref{GresultCC1}, \ref{GresultCC2}, \ref{GresultCC3}, \ref{GresultCC4}, \ref{GresultCC5}, \ref{GresultCC6}, \ref{GresultCC7}, \ref{GresultCC8}, \ref{GresultCC9}, \ref{GresultCC10}, \ref{GresultCC11} and~\ref{GresultCC12}. Therein, the considered meson-meson, diquark-antidiquark and K-type configurations are listed in the first column; if possible, the experimental value of the non-interacting di-meson threshold is labeled in parentheses. In the second column, each channel is assigned with an index, which indicates a particular combination of spin ($\chi_J^{\sigma_i}$), flavor ($\chi_I^{f_j}$) and color ($\chi_k^c$) wave functions, that are shown explicitly in the third column. The theoretical mass calculated in each channel is shown in the fourth column, and the coupled result for each kind of configuration is presented in the last one. Last row of the table indicates the lowest-lying mass, which is obtained in a complete coupled-channel calculation within the real-range formalism.

In a further step, a complete coupled-channels calculation is performed using the CSM in each $I(J^P)$ $\bar{q}q\bar{s}Q$ tetraquark system. Figs.~\ref{PP1} to~\ref{PP12} show the distribution of complex eigenenergies, and therein, the obtained resonance states are indicated inside circles. Several insights about the nature of these resonances are given by calculating their interquark sizes and dominant components; correspondingly, results are listed among Tables~\ref{GresultR1}, \ref{GresultR2}, \ref{GresultR3}, \ref{GresultR4}, \ref{GresultR5}, \ref{GresultR6}, \ref{GresultR7}, \ref{GresultR8}, \ref{GresultR9}, \ref{GresultR10}, \ref{GresultR11} and~\ref{GresultR12}. Particularly, since the $SU(3)$ flavor symmetry is considered for the $\bar{q}q\bar{s}Q$ tetraquark systems, four kinds of quark distances, which are $r_{q\bar{q}}$, $r_{\bar{q}\bar{q}}$, $r_{c\bar{q}}$ and $r_{qc}$ $(q=u,\,d,\,s)$, are calculated. Finally, a summary of our most salient results is presented in Table~\ref{GresultCCT}.

Now let us proceed to describe in detail our theoretical findings for each sector of $\bar{q}q\bar{s}Q$ tetraquarks.

\subsection{The $\mathbf{\bar{q}q\bar{s}c \,(q=u,\,d)}$ tetraquarks}

Several resonances whose masses range from $2.8$ GeV to $3.5$ GeV are obtained in this tetraquark sector. Particularly, the experimentally reported exotic hadron $T_{c\bar{s}}(2900)$~\cite{LHCb:2022sfr, LHCb:2022lzp} can be well identified within the $1(0^+)$ state. Each iso-scalar and -vector sectors with total spin and parity $J^P=0^+$, $1^+$ and $2^+$ shall be discussed individually below.


\begin{table}[!t]
\caption{\label{GresultCC1} Lowest-lying $\bar{q}q\bar{s}c$ tetraquark states with $I(J^P)=0(0^+)$ calculated within the real range formulation of the chiral quark model.
The allowed meson-meson, diquark-antidiquark and K-type configurations are listed in the first column; when possible, the experimental value of the non-interacting meson-meson threshold is labeled in parentheses. Each channel is assigned an index in the 2nd column, it reflects a particular combination of spin ($\chi_J^{\sigma_i}$), flavor ($\chi_I^{f_j}$) and color ($\chi_k^c$) wave functions that are shown explicitly in the 3rd column. The theoretical mass obtained in each channel is shown in the 4th column and the coupled result for each kind of configuration is presented in the 5th column.
When a complete coupled-channels calculation is performed, last row of the table indicates the calculated lowest-lying mass (unit: MeV).}
\begin{ruledtabular}
\begin{tabular}{lcccc}
~~Channel   & Index & $\chi_J^{\sigma_i}$;~$\chi_I^{f_j}$;~$\chi_k^c$ & $M$ & Mixed~~ \\
        &   &$[i; ~j; ~k]$ &  \\[2ex]
$(\eta D_s)^1 (2516)$          & 1  & [1;~1;~1]  & $2678$ & \\
$(\omega D^*_s)^1 (2894)$  & 2  & [2;~1;~1]   & $2811$ &  \\
$(K D)^1 (2364)$          & 3  & [1;~1;~1]  & $2378$ & \\
$(K^* D^*)^1 (2899)$  & 4  & [2;~1;~1]   & $2924$ & $2378$ \\[2ex]
$(\eta D_s)^8$          & 5  & [1;~1;~2]  & $3293$ & \\
$(\omega D^*_s)^8$  & 6  & [2;~1;~2]   & $3107$ &  \\
$(K D)^8$          & 7  & [1;~1;~2]  & $3169$ & \\
$(K^* D^*)^8$  & 8  & [2;~1;~2]   & $3186$ & $2932$ \\[2ex]
$(qc)(\bar{q}\bar{s})$      & 9   & [3;~2;~4]  & $3188$ & \\
$(qc)(\bar{q}\bar{s})$      & 10   & [3;~3;~3]  & $2913$ & \\
$(qc)^*(\bar{q}\bar{s})^*$  & 11  & [4;~2;~3]   & $3186$ & \\
$(qc)^*(\bar{q}\bar{s})^*$  & 12  & [4;~3;~4]   & $3074$ & $2777$ \\[2ex]
$K_1$  & 13  & [5;~1;~5]   & $3102$ & \\
  & 14  & [6;~1;~5]   & $3286$ & \\
  & 15  & [5;~1;~6]   & $3066$ & \\
  & 16  & [6;~1;~6]   & $3071$ & $2961$ \\[2ex]
$K_2$  & 17  & [7;~1;~7]   & $3094$ & \\
  & 18  & [8;~1;~7]   & $3228$ & \\
  & 19  & [7;~1;~8]   & $3207$ & \\
  & 20  & [8;~1;~8]   & $3080$ & $2850$ \\[2ex]
$K_3$  & 21  & [9;~2;~10]   & $2920$ & \\
  & 22  & [9;~3;~9]   & $3071$ & \\
  & 23  & [10;~2;~9]   & $3190$ & \\
  & 24  & [10;~3;~10]   & $2920$ & $2723$ \\[2ex]
\multicolumn{4}{c}{Complete coupled-channels:} & $2378$
\end{tabular}
\end{ruledtabular}
\end{table}

\begin{figure}[!t]
\includegraphics[clip, trim={3.0cm 1.9cm 3.0cm 1.0cm}, width=0.45\textwidth]{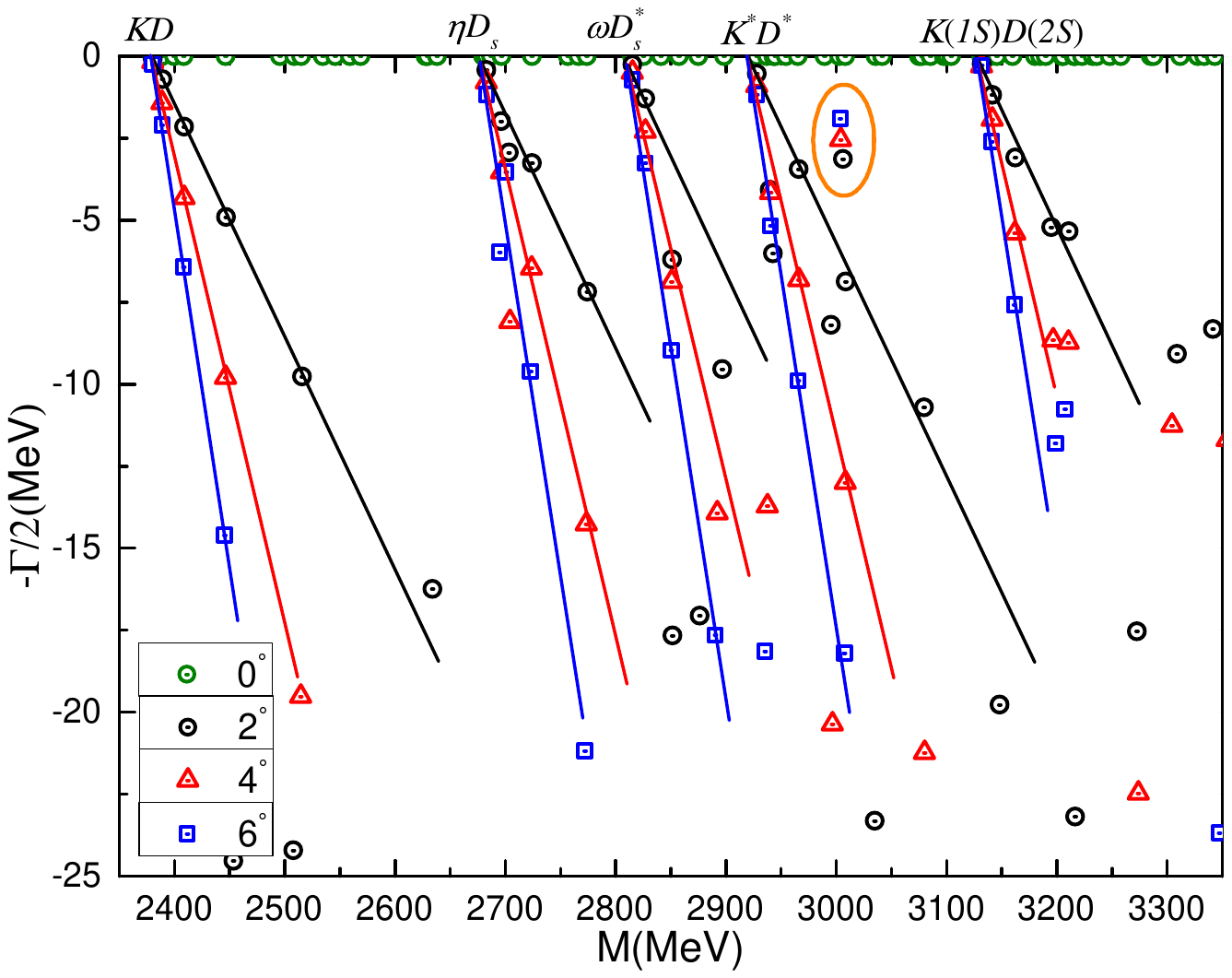}
\caption{\label{PP1} The complete coupled-channel calculation of $\bar{q}q\bar{s}c$ tetraquark system with $I(J^P)=0(0^+)$ quantum numbers.}
\end{figure}

\begin{table}[!t]
\caption{\label{GresultR1} Compositeness of the exotic resonance obtained in a complete coupled-channel calculation in the $0(0^+)$ state of $\bar{q}q\bar{s}c$ tetraquark. Particularly, the first column is the resonance pole labeled by $M+i\Gamma$, unit in MeV; the second one is the distance between any two quarks or quark-antiquark ($q=u,d,s$), unit in fm; and the component of resonance state ($S$: dimeson structure in color-singlet channel; $H$: dimeson structure in hidden-color channel; $Di$: diquark-antiquark configuration; $K$: K-type configuration).}
\begin{ruledtabular}
\begin{tabular}{lccc}
Resonance       & \multicolumn{3}{c}{Structure} \\[2ex]
$3006+i6.3$   & \multicolumn{3}{c}{$r_{q\bar{q}}:1.64$;\,\,\,\,\,$r_{\bar{q}\bar{q}}:2.23$;\,\,\,\,\,$r_{c\bar{q}}:1.70$;\,\,\,\,\,$r_{qc}:2.19$} \\
& \multicolumn{3}{c}{$S$: 21.3\%;\, $H$: 9.1\%;\, $Di$: 13.1\%;\, $K$: 56.5\%}
\end{tabular}
\end{ruledtabular}
\end{table}

{\bf The $\bm{I(J^P)=0(0^+)}$ sector:} Four meson-meson configurations, $\eta D_s$, $\omega D^*_s$, $K D$ and $K^* D^*$ in both color-singlet and -octet channels, four diquark-antidiquark structures, along with three K-type configurations are individually calculated in Table~\ref{GresultCC1}. The lowest channel is the color-singlet state of $K D$, whose theoretical mass is $2378$ MeV, the other three meson-meson configurations with the same color channel are in an energy region from $2.6$ to $3.0$ GeV. No bound state is found. Additionally, the single channel calculations are also performed in each exotic structures. The hidden-color channels of di-meson configurations are generally located in the $3.1-3.3$ GeV interval. This result also holds for the diquark-antidiquark and K-type structures, although the lowest masses of a $(qc)(\bar{q}\bar{s})$ and a $K_3$-type channels are $2.9$ GeV.
 
After partially coupled-channel computations are performed in six configurations listed in Table~\ref{GresultCC1}, only the scattering state of $K D$ and several color resonances, which masses are in an energy region of $2.7-2.9$ GeV, are obtained. Meanwhile, the lowest-lying mass of $2378$ MeV for a $K D$ scattering state remains even in the complete coupled-channel calculation.
 
Figure~\ref{PP1} presents the distribution of complex energies for the $\bar{q}q\bar{s}c$ tetraquark in the $0(0^+)$ channel calculated by the CSM in a fully coupled-channels investigation. Particularly, within a mass region from $2.35$ to $3.35$ GeV, five scattering states that include ground states of $K D$, $\eta D_s$, $\omega D^*_s$, $K^* D^*$ and the radial excitation of $K(1S) D(2S)$ are well presented. However, apart from the vast majority of scattering dots, one stable resonance pole is found and circled. The mass and width is $3006$ and $6.3$ MeV, respectively.
 
Table~\ref{GresultR1} shows the compositeness of the resonance state. Firstly, it is a loosely-bound structure with the quark-antiquark distance $\sim 1.6$ fm and $\sim 2.2$ fm for the $qc$ and $\bar{q}\bar{q}$ pairs. Besides, there is a strong coupling among singlet-, hidden-color, diquark-antidiquark and K-type channels. The golden decays for this resonance are $\omega D^*_s$ and $K^* D^*$, which are the dominant components $(21.3\%)$ of the color-singlet channels.


\begin{table}[!t]
\caption{\label{GresultCC2} Lowest-lying $\bar{q}q\bar{s}c$ tetraquark states with $I(J^P)=0(1^+)$ calculated within the real range formulation of the chiral quark model. Results are similarly organized as those in Table~\ref{GresultCC1} (unit: MeV).}
\begin{ruledtabular}
\begin{tabular}{lcccc}
~~Channel   & Index & $\chi_J^{\sigma_i}$;~$\chi_I^{f_j}$;~$\chi_k^c$ & $M$ & Mixed~~ \\
        &   &$[i; ~j; ~k]$ &  \\[2ex]
$(\eta D^*_s)^1 (2660)$          & 1  & [1;~1;~1]  & $2804$ & \\
$(\omega D_s)^1 (2750)$  & 2  & [2;~1;~1]   & $2685$ &  \\
$(\omega D^*_s)^1 (2894)$  & 3  & [3;~1;~1]   & $2811$ &  \\
$(K D^*)^1 (2501)$       & 4  & [1;~1;~1]   & $2498$ &  \\
$(K^* D)^1 (2762)$      & 5  & [2;~1;~1]  & $2804$ & \\
$(K^* D^*)^1 (2899)$  & 6  & [3;~1;~1]   & $2924$ & $2498$ \\[2ex]
$(\eta D^*_s)^8$  & 7  & [1;~1;~2]  & $3296$ & \\
$(\omega D_s)^8$    & 8  & [2;~1;~2]   & $3179$ &  \\
$(\omega D^*_s)^8$  & 9  & [3;~1;~2]   & $3147$ &  \\
$(K D^*)^8$      & 10  & [1;~1;~2]   & $3174$ &  \\
$(K^* D)^8$      & 11 & [2;~1;~2]  & $3191$ & \\
$(K^* D^*)^8$      & 12  & [3;~1;~2]   & $3192$ & $2897$ \\[2ex]
$(qc)(\bar{q}\bar{s})^*$      & 13   & [4;~2;~4]  & $3181$ & \\
$(qc)(\bar{q}\bar{s})^*$      & 14   & [4;~3;~3]  & $2949$ & \\
$(qc)^*(\bar{q}\bar{s})$      & 15   & [5;~2;~3]  & $3159$ & \\
$(qc)^*(\bar{q}\bar{s})$      & 16  & [5;~3;~4]   & $3162$ & \\
$(qc)^*(\bar{q}\bar{s})^*$  & 17  & [6;~2;~3]   & $3169$ &  \\
$(qc)^*(\bar{q}\bar{s})^*$  & 18  & [6;~3;~4]   & $3094$ & $2852$ \\[2ex]
$K_1$  & 19  & [7;~1;~5]   & $3178$ & \\
  & 20  & [8;~1;~5]   & $3121$ & \\
  & 21  & [9;~1;~5]   & $3291$ & \\
  & 22  & [7;~1;~6]   & $3044$ & \\
  & 23  & [8;~1;~6]   & $3053$ & \\
  & 24  & [9;~1;~6]   & $3113$ & $3005$ \\[2ex]
$K_2$  & 25  & [10;~1;~7]   & $3109$ & \\
  & 26  & [11;~1;~7]   & $3186$ & \\
  & 27  & [12;~1;~7]   & $3187$ & \\
  & 28  & [10;~1;~8]   & $3045$ & \\
  & 29  & [11;~1;~8]   & $3213$ & \\
  & 30  & [12;~1;~8]   & $3187$ & $2956$ \\[2ex]
$K_3$  & 31  & [13;~2;~10]   & $3085$ & \\
  & 32  & [13;~3;~9]   & $3083$ & \\
  & 33  & [14;~2;~10]   & $3196$ & \\
  & 34  & [14;~3;~9]   & $3112$ & \\
  & 35  & [15;~2;~10]   & $3178$ & \\
  & 36  & [15;~3;~9]   & $2946$ & $2787$ \\[2ex]  
\multicolumn{4}{c}{Complete coupled-channels:} & $2498$
\end{tabular}
\end{ruledtabular}
\end{table}

\begin{figure}[!t]
\includegraphics[clip, trim={3.0cm 1.9cm 3.0cm 1.0cm}, width=0.45\textwidth]{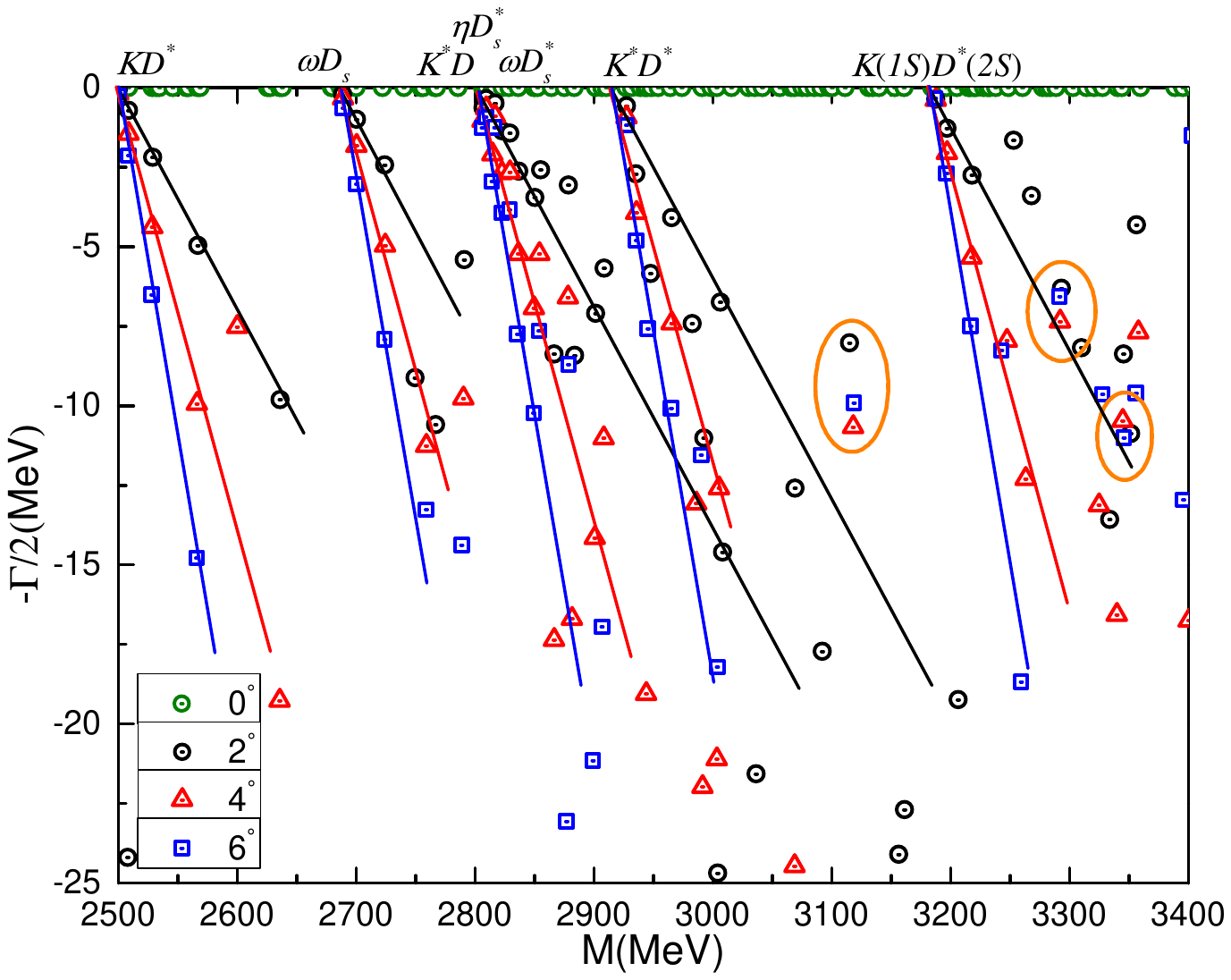}
\caption{\label{PP2} The complete coupled-channel calculation of $\bar{q}q\bar{s}c$ tetraquark system with $I(J^P)=0(1^+)$.}
\end{figure}

\begin{table}[!t]
\caption{\label{GresultR2} Compositeness of exotic resonances obtained in a complete coupled-channel calculation in the $0(1^+)$ state of $\bar{q}q\bar{s}c$ tetraquark. Results are similarly organized as those in Table~\ref{GresultR1}.}
\begin{ruledtabular}
\begin{tabular}{lccc}
Resonance       & \multicolumn{3}{c}{Structure} \\[2ex]
$3119+i19.8$   & \multicolumn{3}{c}{$r_{q\bar{q}}:1.41$;\,\,\,\,\,$r_{\bar{q}\bar{q}}:1.63$;\,\,\,\,\,$r_{c\bar{q}}:1.03$;\,\,\,\,\,$r_{qc}:1.61$} \\
& \multicolumn{3}{c}{$S$: 8.5\%;\, $H$: 18.7\%;\, $Di$: 29.6\%;\, $K$: 43.2\%}\\[1.5ex]
$3292+i13.1$   & \multicolumn{3}{c}{$r_{q\bar{q}}:1.14$;\,\,\,\,\,$r_{\bar{q}\bar{q}}:1.13$;\,\,\,\,\,$r_{c\bar{q}}:1.14$;\,\,\,\,\,$r_{qc}:1.37$} \\
& \multicolumn{3}{c}{$S$: 9.6\%;\, $H$: 13.9\%;\, $Di$: 34.5\%;\, $K$: 42.0\%}\\[1.5ex]
$3346+i22.0$   & \multicolumn{3}{c}{$r_{q\bar{q}}:1.54$;\,\,\,\,\,$r_{\bar{q}\bar{q}}:1.75$;\,\,\,\,\,$r_{c\bar{q}}:1.49$;\,\,\,\,\,$r_{qc}:1.64$} \\
& \multicolumn{3}{c}{$S$: 6.7\%;\, $H$: 8.8\%;\, $Di$: 34.7\%;\, $K$: 49.8\%}
\end{tabular}
\end{ruledtabular}
\end{table}

{\bf The $\bm{I(J^P)=0(1^+)}$ sector:} 36 channels contribute to this case, and results in real-range calculations are listed in Table~\ref{GresultCC2}. Firstly, the lowest mass, $2498$ MeV, in a single channel computation is just the theoretical threshold value of $K D^*$. The other dimeson channels, which include the $\eta D^*_s$, $\omega D^{(*)}_s$ and $K^* D^{(*)}$, are generally located in $2.7-2.9$ GeV. All of these states are of scattering nature. There are also six channels included in each exotic configuration. By referencing the calculated data on each channel, one can find that the lowest masses in the hidden-color, diquark-antidiquark and K-type configurations are all within $3.0-3.3$ GeV. Furthermore, color resonances with structures of diquark-antidiquark and $K_3$-type are still obtained at around $2.95$ GeV.

In a further step, the lowest coupled-channel masses within each considered configuration are $2.50$, $2.90$, $2.85$, $3.00$, $2.96$ and $2.79$ GeV, respectively. These results indicate that the coupling effect is quite weak in color-singlet channels, but it becomes stronger in other configurations. However, the bound state is still unavailable even in a complete coupled-channel situation.

In order to find possible resonance state in an excited energy region of $2.5-3.4$ GeV, the fully coupled-channel calculation is further performed by the CSM, and results are plotted in a complex energy plane of Fig.~\ref{PP2}. Therein, seven meson-meson scattering states are generally presented. They are ground states of $K^{(*)} D^{(*)}$, $\omega D^{(*)}_s$ and $\eta D^*_s$, and the first radial excited state of $K(1S) D^*(2S)$. However, three stable poles are obtained within the radial excited energy region, and their complex energies read $3119+i19.8$, $3292+i13.1$ and $3346+i22$ MeV, respectively.

Table~\ref{GresultR2} shows particular features of the three resonances. Firstly, the dominant components of them are all of exotic color structure, $i. e.$, the hidden-color, diquark-antidiquark and K-type configurations. Besides, the coupling among these three sectors is strong. The color resonances are also confirmed by calculating their sizes, with internal quark distances of about $1.1-1.7$ fm. These resonances are expected to be experimentally studied in the $K^{(*)} D^{(*)}$ golden decay channels.


\begin{table}[!t]
\caption{\label{GresultCC3} Lowest-lying $\bar{q}q\bar{s}c$ tetraquark states with $I(J^P)=0(2^+)$ calculated within the real range formulation of the chiral quark model. Results are similarly organized as those in Table~\ref{GresultCC1} (unit: MeV).}
\begin{ruledtabular}
\begin{tabular}{lcccc}
~~Channel   & Index & $\chi_J^{\sigma_i}$;~$\chi_I^{f_j}$;~$\chi_k^c$ & $M$ & Mixed~~ \\
        &   &$[i; ~j; ~k]$ &  \\[2ex]
$(\omega D^*_s)^1 (2894)$  & 1  & [1;~1;~1]   & $2811$ &  \\
$(K^* D^*)^1 (2899)$  & 2  & [1;~1;~1]   & $2924$ & $2811$ \\[2ex]
$(\omega D^*_s)^8$  & 3  & [1;~1;~2]   & $3216$ &  \\
$(K^* D^*)^8$  & 4  & [1;~1;~2]   & $3201$ & $3104$ \\[2ex]
$(qc)^*(\bar{q}\bar{s})^*$  & 5  & [1;~2;~3]   & $3132$ & \\
$(qc)^*(\bar{q}\bar{s})^*$  & 6  & [1;~3;~4]   & $3130$ & $3119$ \\[2ex]
$K_1$  & 7  & [1;~1;~5]   & $3200$ & \\
  & 8  & [1;~1;~6]   & $3071$ & $3070$ \\[2ex]
$K_2$  & 9  & [1;~1;~7]   & $3156$ & \\
  & 10  & [1;~1;~8]   & $3159$ & $3115$ \\[2ex]
$K_3$  & 11  & [1;~2;~10]   & $3119$ & \\
  & 12  & [1;~3;~9]   & $3111$ & $3104$ \\[2ex]
\multicolumn{4}{c}{Complete coupled-channels:} & $2811$
\end{tabular}
\end{ruledtabular}
\end{table}

\begin{figure}[!t]
\includegraphics[width=0.45\textwidth, trim={2.3cm 2.0cm 3.0cm 1.0cm}]{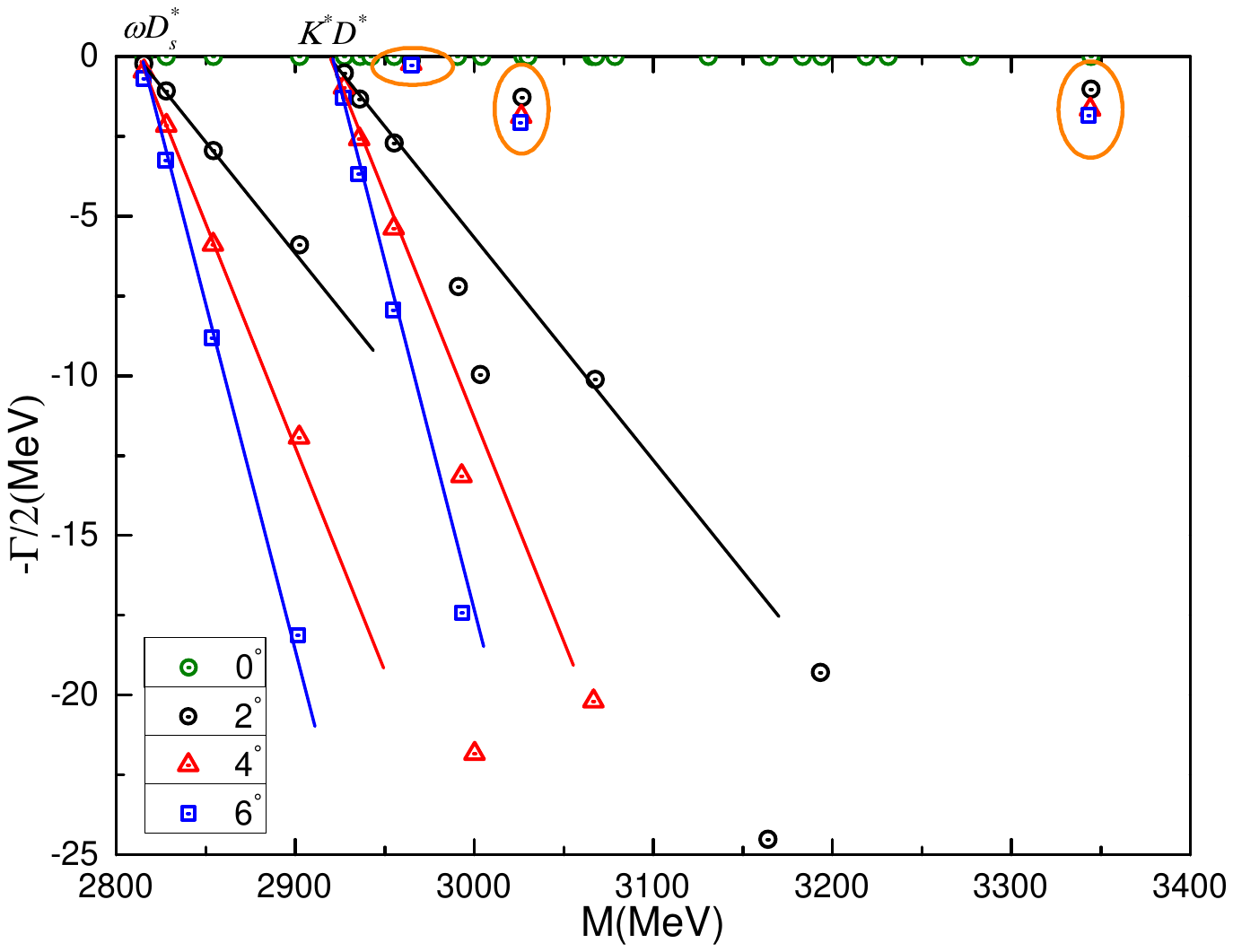}
\caption{\label{PP3} The complete coupled-channel calculation of $\bar{q}q\bar{s}c$ tetraquark system with $I(J^P)=0(2^+)$ quantum numbers.}
\end{figure}

\begin{table}[!t]
\caption{\label{GresultR3} Compositeness of exotic resonances obtained in a complete coupled-channel calculation in the $0(2^+)$ state of $\bar{q}q\bar{s}c$ tetraquark. Results are similarly organized as those in Table~\ref{GresultR1}.}
\begin{ruledtabular}
\begin{tabular}{lccc}
Resonance       & \multicolumn{3}{c}{Structure} \\[2ex]
$2965+i0.5$   & \multicolumn{3}{c}{$r_{q\bar{q}}:1.18$;\,\,\,\,\,$r_{\bar{q}\bar{q}}:1.19$;\,\,\,\,\,$r_{c\bar{q}}:0.86$;\,\,\,\,\,$r_{qc}:1.11$} \\
& \multicolumn{3}{c}{$S$: 18.9\%;\, $H$: 8.1\%;\, $Di$: 27.8\%;\, $K$: 45.2\%}\\[1.5ex]
$3026+i3.8$   & \multicolumn{3}{c}{$r_{q\bar{q}}:1.40$;\,\,\,\,\,$r_{\bar{q}\bar{q}}:1.50$;\,\,\,\,\,$r_{c\bar{q}}:1.02$;\,\,\,\,\,$r_{qc}:1.43$} \\
& \multicolumn{3}{c}{$S$: 15.4\%;\, $H$: 12.6\%;\, $Di$: 7.3\%;\, $K$: 64.7\%}\\[1.5ex]
$3344+i3.3$   & \multicolumn{3}{c}{$r_{q\bar{q}}:1.46$;\,\,\,\,\,$r_{\bar{q}\bar{q}}:1.54$;\,\,\,\,\,$r_{c\bar{q}}:1.38$;\,\,\,\,\,$r_{qc}:1.46$} \\
& \multicolumn{3}{c}{$S$: 13.4\%;\, $H$: 20.2\%;\, $Di$: 18.5\%;\, $K$: 47.9\%}
\end{tabular}
\end{ruledtabular}
\end{table}

{\bf The $\bm{I(J^P)=0(2^+)}$ state:} Two meson-meson channels, $\omega D^*_s$ and $K^* D^*$, should be considered in the highest spin state. In Table~\ref{GresultCC3} one can find that the lowest masses of them are just theoretical threshold values, hence no bound state is found. Meanwhile, other ten channels of exotic structures are generally located in a mass region $3.1-3.2$ GeV. When coupled-channel calculations are performed in each specific configuration, color resonances are located at $\sim 3.1$ GeV, and the scattering state of $\omega D^*_s$, which is the lowest-lying channel, remains at $2.8$ GeV. This extremely weak coupling effect also holds for the complete coupled-channel study.

Nevertheless, three narrow resonances are obtained in a complex analysis on the fully coupled-channels computation. Figure~\ref{PP3} shows the two scattering states of $\omega D^*_s$ and $K^* D^*$ within $2.8-3.4$ GeV. Moreover, there are three stable poles against the descending cut lines when the rotated angle is varied from $0^\circ$ to $6^\circ$.

Naturalness of these narrow resonances can be guessed from Table~\ref{GresultR3}. Particularly, complex energies of resonances are $2965+i0.5$, $3026+i3.8$ and $3344+i3.3$ MeV, respectively. Compact structures are dominant when referring to their interquark sizes, which are about $1.4$ fm. Furthermore, there are strong couplings among the color-singlet, hidden-color, diquark-antidiquark and K-type configurations of these states. Since the singlet-color component are $(\sim 7\%)$ of $\omega D^*_s$ and $K^* D^*$ for these resonances, they can be experimentally investigated in any of the mentioned two-body strong decay process.


\begin{table}[!t]
\caption{\label{GresultCC4} Lowest-lying $\bar{q}q\bar{s}c$ tetraquark states with $I(J^P)=1(0^+)$ calculated within the real range formulation of the chiral quark model. Results are similarly organized as those in Table~\ref{GresultCC1} (unit: MeV).}
\begin{ruledtabular}
\begin{tabular}{lcccc}
~~Channel   & Index & $\chi_J^{\sigma_i}$;~$\chi_I^{f_j}$;~$\chi_k^c$ & $M$ & Mixed~~ \\
        &   &$[i; ~j; ~k]$ &  \\[2ex]
$(\pi D_s)^1 (2108)$          & 1  & [1;~1;~1]  & $2138$ & \\
$(\rho D^*_s)^1 (2882)$  & 2  & [2;~1;~1]   & $2887$ &  \\
$(K D)^1 (2364)$          & 3  & [1;~1;~1]  & $2378$ & \\
$(K^* D^*)^1 (2899)$  & 4  & [2;~1;~1]   & $2924$ & $2138$ \\[2ex]
$(\pi D_s)^8$          & 5  & [1;~1;~2]  & $3177$ & \\
$(\rho D^*_s)^8$  & 6  & [2;~1;~2]   & $3167$ &  \\
$(K D)^8$          & 7  & [1;~1;~2]  & $3169$ & \\
$(K^* D^*)^8$  & 8  & [2;~1;~2]   & $3105$ & $2894$ \\[2ex]
$(qc)(\bar{q}\bar{s})$      & 9   & [3;~2;~4]  & $3188$ & \\
$(qc)(\bar{q}\bar{s})$      & 10   & [3;~3;~3]  & $2913$ & \\
$(qc)^*(\bar{q}\bar{s})^*$  & 11  & [4;~2;~3]   & $3084$ & \\
$(qc)^*(\bar{q}\bar{s})^*$  & 12  & [4;~3;~4]   & $2956$ & $2772$ \\[2ex]
$K_1$  & 13  & [5;~1;~5]   & $3166$ & \\
  & 14  & [6;~1;~5]   & $3167$ & \\
  & 15  & [5;~1;~6]   & $3157$ & \\
  & 16  & [6;~1;~6]   & $2475$ & $2460$ \\[2ex]
$K_2$  & 17  & [7;~1;~7]   & $2972$ & \\
  & 18  & [8;~1;~7]   & $3228$ & \\
  & 19  & [7;~1;~8]   & $3105$ & \\
  & 20  & [8;~1;~8]  & $3080$ & $2826$ \\[2ex]
$K_3$  & 21  & [9;~2;~10]   & $3062$ & \\
  & 22  & [9;~3;~9]   & $2891$ & \\
  & 23  & [10;~2;~9]   & $3190$ & \\
  & 24  & [10;~3;~10]   & $2920$ & $2700$ \\[2ex]
\multicolumn{4}{c}{Complete coupled-channels:} & $2138$
\end{tabular}
\end{ruledtabular}
\end{table}

\begin{figure}[!t]
\includegraphics[clip, trim={3.0cm 1.9cm 3.0cm 1.0cm}, width=0.45\textwidth]{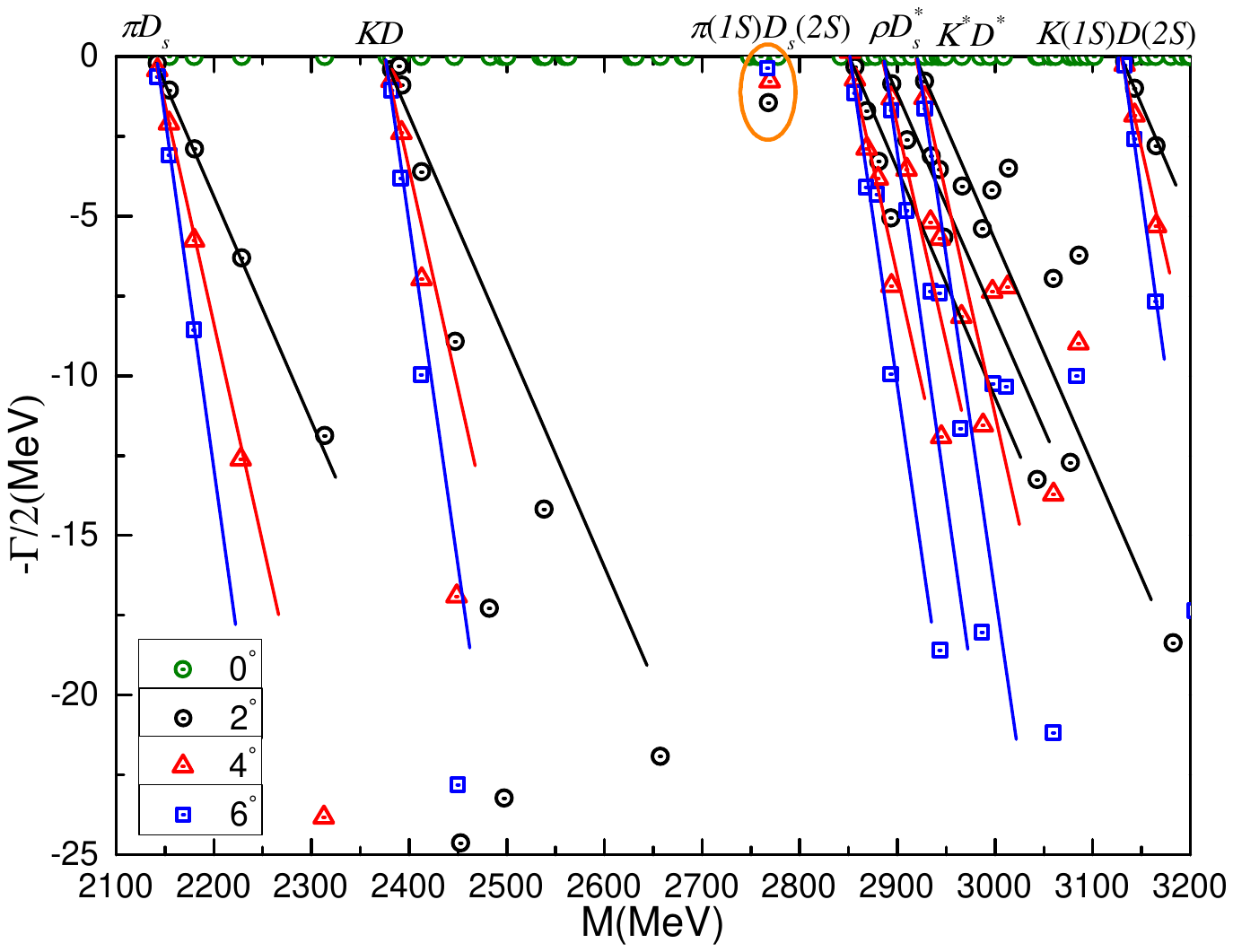}
\caption{\label{PP4} The complete coupled-channels calculation of $\bar{q}q\bar{s}c$ tetraquark system with $I(J^P)=1(0^+)$ quantum numbers.}
\end{figure}

\begin{table}[!t]
\caption{\label{GresultR4} Compositeness of the exotic resonance obtained in a complete coupled-channel calculation in the $1(0^+)$ state of $\bar{q}q\bar{s}c$ tetraquark. Results are similarly organized as those in Table~\ref{GresultR1}.}
\begin{ruledtabular}
\begin{tabular}{lccc}
Resonance       & \multicolumn{3}{c}{Structure} \\[2ex]
$2770+i1.5$   & \multicolumn{3}{c}{$r_{q\bar{q}}:1.05$;\,\,\,\,\,$r_{\bar{q}\bar{q}}:1.80$;\,\,\,\,\,$r_{c\bar{q}}:1.59$;\,\,\,\,\,$r_{qc}:1.75$} \\
& \multicolumn{3}{c}{$S$: 15.6\%;\, $H$: 19.2\%;\, $Di$: 30.3\%;\, $K$: 34.9\%}
\end{tabular}
\end{ruledtabular}
\end{table}

{\bf The $\bm{I(J^P)=1(0^+)}$ sector:} This case is similar to the $0(0^+)$ channel, \emph{i.e.} 24 channels are investigated as shown in Table~\ref{GresultCC4}. Firstly, $\pi D_s$, $\rho D^*_s$, $K D$ and $K^* D^*$ channels in both singlet- and hidden-color configurations are calculated. The lowest-lying state is the $\pi D_s$ scattering state with the theoretical threshold value of $2138$ MeV. Besides, the other three meson-meson structures in color-singlet channels are also unbound, and the four hidden-color channels are generally located at $3.17$ GeV. Concerning the single channel computations of diquark-antidiquark and K-type configurations, the lowest masses are generally distributed within an energy region of $2.9-3.2$ GeV, except for a $K_1$-type channel with mass at $2475$ MeV.

We do not find bound states when partially and fully coupled-channels calculations in the real-range approximation are performed; the lowest-lying $\pi D_s$ scattering state remains at $2138$ MeV. Coupled masses in other exotic structures are located at $\sim 2.8$ GeV, apart from the $2.46$ GeV of $K_1$ channels. Additionally, a resonance state at $2.8$ GeV, which is compatible with the experimentally observed $T_{c\bar{s}}(2900)$ state~\cite{LHCb:2022sfr, LHCb:2022lzp}, is obtained in the complete coupled-channel study in the complex range formulation. In particular, within the $2.1-3.2$ GeV energy region of Fig.~\ref{PP4}, six continuum states of $\pi D_s$, $K^{(*)} D^{(*)}$, $\rho D^*_s$, $\pi(1S) D_s(2S)$ and $K(1S) D(2S)$ are clearly presented. However, one stable resonance pole is circled, and the complex energy is predicted to be $2770+i1.5$ MeV.

Table~\ref{GresultR4} lists the interquark distances and wavefunction components of the predicted exotic resonance. Apparently, its size is around $1.6$ fm, and there is a strong coupling among the four considered configurations, \emph{i.e.}, the color-singlet, hidden-color, diquark-antidiquark and K-type structures. The color-singlet channels of $\pi D_s$ and $K D$ are comparable ($\sim 8\%$), hence they are suggested to be the golden decay channels.


\begin{table}[!t]
\caption{\label{GresultCC5} Lowest-lying $\bar{q}q\bar{s}c$ tetraquark states with $I(J^P)=1(1^+)$ calculated within the real range formulation of the chiral quark model. Results are similarly organized as those in Table~\ref{GresultCC1} (unit: MeV).}
\begin{ruledtabular}
\begin{tabular}{lcccc}
~~Channel   & Index & $\chi_J^{\sigma_i}$;~$\chi_I^{f_j}$;~$\chi_k^c$ & $M$ & Mixed~~ \\
        &   &$[i; ~j; ~k]$ &  \\[2ex]
$(\pi D^*_s)^1 (2252)$          & 1  & [1;~1;~1]  & $2264$ & \\
$(\rho D_s)^1 (2738)$  & 2  & [2;~1;~1]   & $2761$ &  \\
$(\rho D^*_s)^1 (2882)$  & 3  & [3;~1;~1]   & $2887$ &  \\
$(K D^*)^1 (2501)$      & 4  & [1;~1;~1]   & $2498$ &  \\
$(K^* D)^1 (2762)$      & 5  & [2;~1;~1]  & $2804$ & \\
$(K^* D^*)^1 (2899)$  & 6  & [3;~1;~1]   & $2924$ & $2264$ \\[2ex]
$(\pi D^*_s)^8$  & 7  & [1;~1;~2]  & $3181$ & \\
$(\rho D_s)^8$    & 8  & [2;~1;~2]   & $3231$ &  \\
$(\rho D^*_s)^8$  & 9  & [3;~1;~2]   & $3203$ &  \\
$(K D^*)^8$      & 10  & [1;~1;~2]   & $3174$ &  \\
$(K^* D)^8$      & 11 & [2;~1;~2]  & $3191$ & \\
$(K^* D^*)^8$     & 12  & [3;~1;~2]   & $3153$ & $2951$ \\[2ex]
$(qc)(\bar{q}\bar{s})^*$      & 13   & [4;~2;~4]  & $3181$ & \\
$(qc)(\bar{q}\bar{s})^*$      & 14   & [4;~3;~3]  & $2949$ & \\
$(qc)^*(\bar{q}\bar{s})$      & 15   & [5;~2;~3]  & $3159$ & \\
$(qc)^*(\bar{q}\bar{s})$      & 16  & [5;~3;~4]   & $3162$ & \\
$(qc)^*(\bar{q}\bar{s})^*$  & 17  & [6;~2;~3]   & $3119$ &  \\
$(qc)^*(\bar{q}\bar{s})^*$  & 18  & [6;~3;~4]   & $3041$ & $2871$ \\[2ex]
$K_1$  & 19  & [7;~1;~5]   & $3233$ & \\
  & 20  & [8;~1;~5]   & $3183$ & \\
  & 21  & [9;~1;~5]   & $3173$ & \\
  & 22  & [7;~1;~6]   & $3134$ & \\
  & 23  & [8;~1;~6]   & $3144$ & \\
  & 24  & [9;~1;~6]   & $2518$ & $2509$ \\[2ex]
$K_2$  & 25  & [10;~1;~7]   & $3158$ & \\
  & 26  & [11;~1;~7]   & $3081$ & \\
  & 27  & [12;~1;~7]   & $3187$ & \\
  & 28  & [10;~1;~8]   & $3105$ & \\
  & 29  & [11;~1;~8]   & $3110$ & \\
  & 30  & [12;~1;~8]   & $3187$ & $2972$ \\[2ex]
$K_3$  & 31  & [13;~2;~10]   & $3149$ & \\
  & 32  & [13;~3;~9]   & $3148$ & \\
  & 33  & [14;~2;~10]   & $3069$ & \\
  & 34  & [14;~3;~9]   & $2942$ & \\
  & 35  & [15;~2;~10]   & $3178$ & \\
  & 36  & [15;~3;~9]   & $2946$ & $2791$ \\[2ex] 
\multicolumn{4}{c}{Complete coupled-channels:} & $2264$
\end{tabular}
\end{ruledtabular}
\end{table}

\begin{figure}[!t]
\includegraphics[clip, trim={3.0cm 1.9cm 3.0cm 1.0cm}, width=0.45\textwidth]{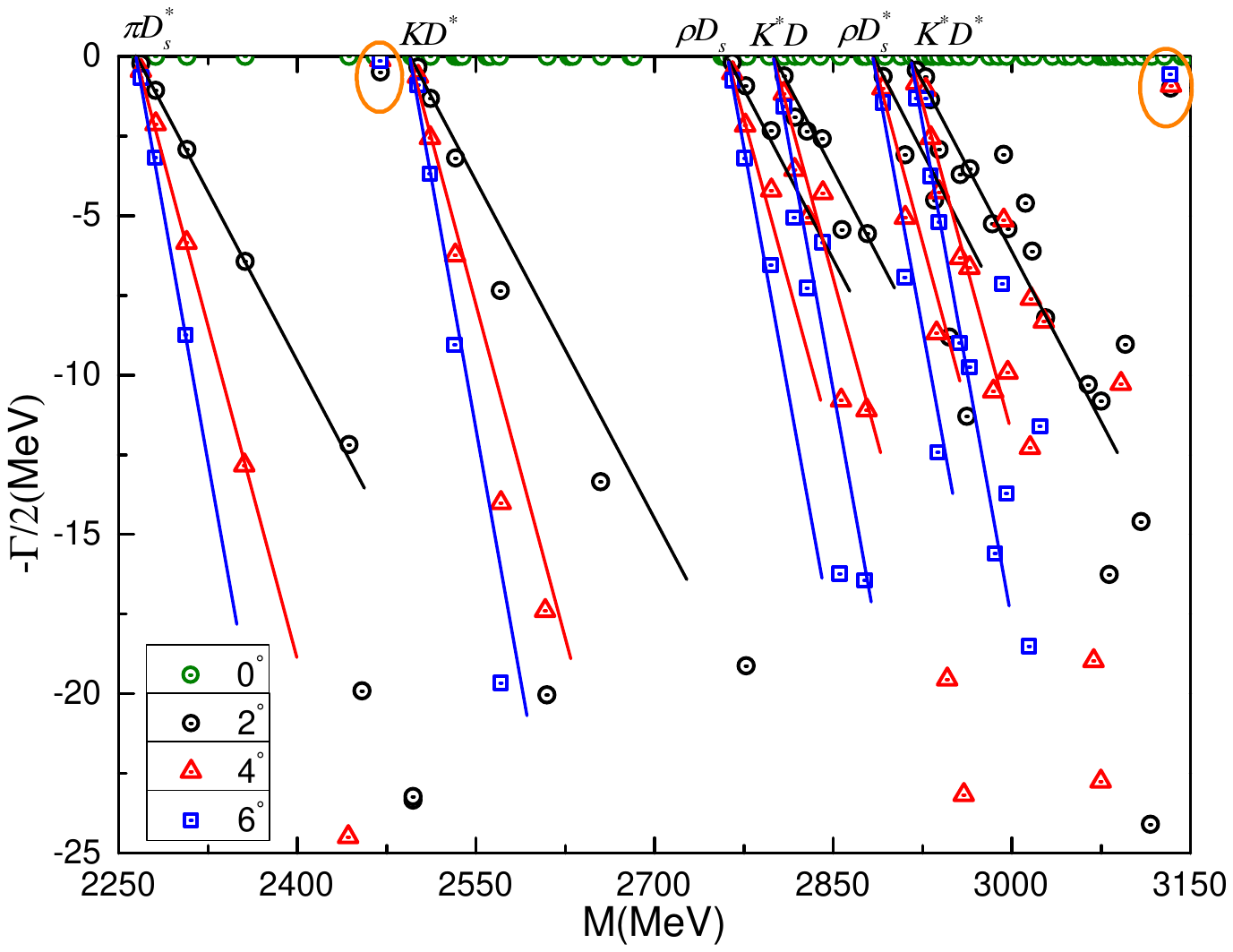} \\[1ex]
\includegraphics[clip, trim={3.0cm 1.9cm 3.0cm 1.0cm}, width=0.45\textwidth]{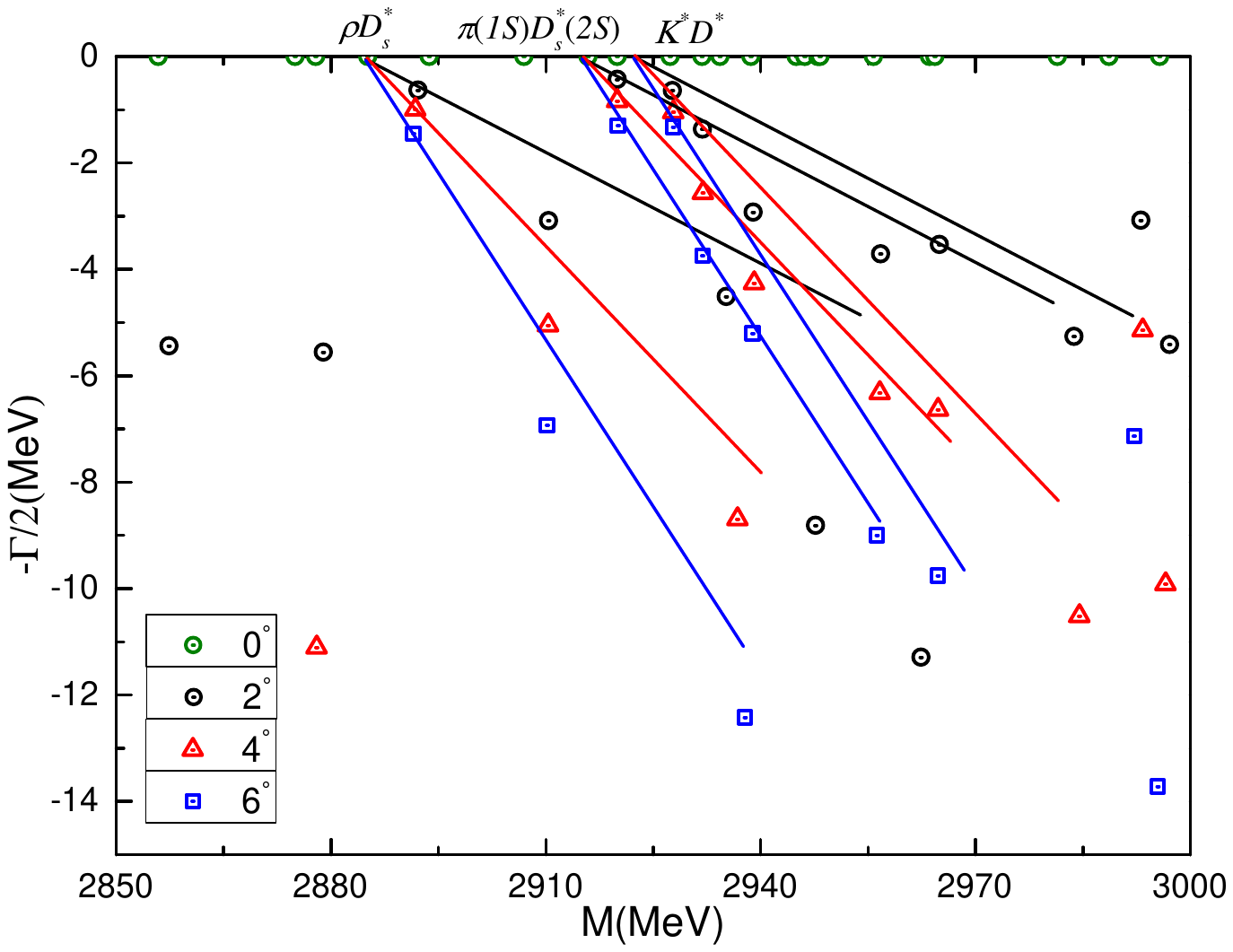}
\caption{\label{PP5} The complete coupled-channels calculation of $\bar{q}q\bar{s}c$ tetraquark system with $I(J^P)=1(1^+)$ quantum numbers. Particularly, the bottom panel is enlarged parts of dense energy region from $2.85\,\text{GeV}$ to $3.00\,\text{GeV}$.}
\end{figure}

\begin{table}[!t]
\caption{\label{GresultR5} Compositeness of exotic resonances obtained in a complete coupled-channel calculation in the $1(1^+)$ state of $\bar{q}q\bar{s}c$ tetraquark. Results are similarly organized as those in Table~\ref{GresultR1}.}
\begin{ruledtabular}
\begin{tabular}{lccc}
Resonance       & \multicolumn{3}{c}{Structure} \\[2ex]
$2470+i1.0$   & \multicolumn{3}{c}{$r_{q\bar{q}}:0.92$;\,\,\,\,\,$r_{\bar{q}\bar{q}}:0.95$;\,\,\,\,\,$r_{c\bar{q}}:0.88$;\,\,\,\,\,$r_{qc}:0.79$} \\
& \multicolumn{3}{c}{$S$: 26.5\%;\, $H$: 4.0\%;\, $Di$: 13.5\%;\, $K$: 56.0\%}\\[1.5ex]
$3134+i2.0$   & \multicolumn{3}{c}{$r_{q\bar{q}}:1.79$;\,\,\,\,\,$r_{\bar{q}\bar{q}}:1.87$;\,\,\,\,\,$r_{c\bar{q}}:1.92$;\,\,\,\,\,$r_{qc}:1.61$} \\
& \multicolumn{3}{c}{$S$: 16.1\%;\, $H$: 18.0\%;\, $Di$: 16.6\%;\, $K$: 49.3\%}
\end{tabular}
\end{ruledtabular}
\end{table}

{\bf The $\bm{I(J^P)=1(1^+)}$ sector:} Among the six channels of the considered $\bar{q}q\bar{s}c$ tetraquark configurations, which include singlet-, hidden- color, diquark-antidiquark and K-type structures, the lowest-lying state is the $\pi D^*_s$ scattering state with the theoretical threshold at $2264$ MeV. Meanwhile, this unbound nature remains unchanged in coupled-channel computations. And the other five di-meson configurations in the color-singlet channels, which are $\rho D^{(*)}_s$ and $K^{(*)} D^{(*)}$, are also unbound. Masses of exotic color channels are generally located in the energy region $2.9-3.2$ GeV, except for a $K_1$-type channel whose calculated mass is $2518$ MeV. When coupled-channels calculations are performed in each of these structures, the lowest masses of hidden-color, diquark-antidiquark, $K_1$, $K_2$ and $K_3$ channels are $2951$, $2871$, $2509$, $2972$ and $2791$ MeV, respectively. Although bound states are unavailable, the mentioned excited states obtained in each exotic color configuration may be good candidates of color resonances for the $\bar{q}q\bar{s}c$ tetraquark system.

Furthermore, Fig.~\ref{PP5} shows the distribution of complex energies in the fully coupled-channels calculation using CSM. In particular, the top panel presents six scattering states, which were discussed above. Within $2.25-3.15$ GeV, two stable poles are obtained and they are indicated within circles. The lower resonance pole is at $2470+i1$ MeV whereas the higher one is at $3134+i2$ MeV. At last, the bottom panel of Fig.~\ref{PP5} is an enlarged part of $2.85-3.00$ GeV. Therein, no stable resonance is found and only three scattering states of $\rho D^*_s$, $\pi(1S) D^*_s(2S)$ and $K^* D^*$ are presented.

Some insight about the nature of the narrow resonances can be found in Table~\ref{GresultR5}. In particular, a compact $\bar{q}q\bar{s}c$ tetraquark structure is predicted for the lower resonance, its size is about $0.9$ fm. The coupling is strong among color-singlet $(27\%)$, diquark-antidiquark $(14\%)$ and K-type $(56\%)$ channels. Besides, the dominant meson-meson decay channel is $\pi D^*_s$, which is expected to be confirmed in future experiments. On the other hand, the higher resonance is a loosely-bound structure with size $\sim 1.8$ fm. Ratios between components are similar to the case of the lower resonance. It is suggested that this state be further studied by high energy experiments in the $\pi D^*_s$ and $K^* D^*$ decay channels.


\begin{table}[!t]
\caption{\label{GresultCC6} Lowest-lying $\bar{q}q\bar{s}c$ tetraquark states with $I(J^P)=1(2^+)$ calculated within the real range formulation of the chiral quark model. Results are similarly organized as those in Table~\ref{GresultCC1} (unit: MeV).}
\begin{ruledtabular}
\begin{tabular}{lcccc}
~~Channel   & Index & $\chi_J^{\sigma_i}$;~$\chi_I^{f_j}$;~$\chi_k^c$ & $M$ & Mixed~~ \\
        &   &$[i; ~j; ~k]$ &  \\[2ex]
$(\rho D^*_s)^1 (2882)$  & 1  & [1;~1;~1]   & $2887$ &  \\
$(K^* D^*)^1 (2899)$  & 2  & [1;~1;~1]   & $2924$ & $2887$ \\[2ex]
$(\rho D^*_s)^8$  & 3  & [1;~1;~2]   & $3265$ &  \\
$(K^* D^*)^8$  & 4  & [1;~1;~2]   & $3235$ & $3140$ \\[2ex]
$(qc)^*(\bar{q}\bar{s})^*$  & 5  & [1;~2;~3]   & $3181$ & \\
$(qc)^*(\bar{q}\bar{s})^*$  & 6  & [1;~3;~4]   & $3174$ & $3166$ \\[2ex]
$K_1$  & 7  & [1;~1;~5]   & $3253$ & \\
  & 8  & [1;~1;~6]   & $3161$ & $3160$ \\[2ex]
$K_2$  & 9  & [1;~1;~7]   & $3202$ & \\
  & 10  & [1;~1;~8]   & $3210$ & $3166$ \\[2ex]
$K_3$  & 11  & [1;~2;~10]   & $3180$ & \\
  & 12  & [1;~3;~9]   & $3174$ & $3167$ \\[2ex]
\multicolumn{4}{c}{Complete coupled-channels:} & $2887$
\end{tabular}
\end{ruledtabular}
\end{table}

\begin{figure}[!t]
\includegraphics[width=0.49\textwidth, trim={2.3cm 2.0cm 2.0cm 1.0cm}]{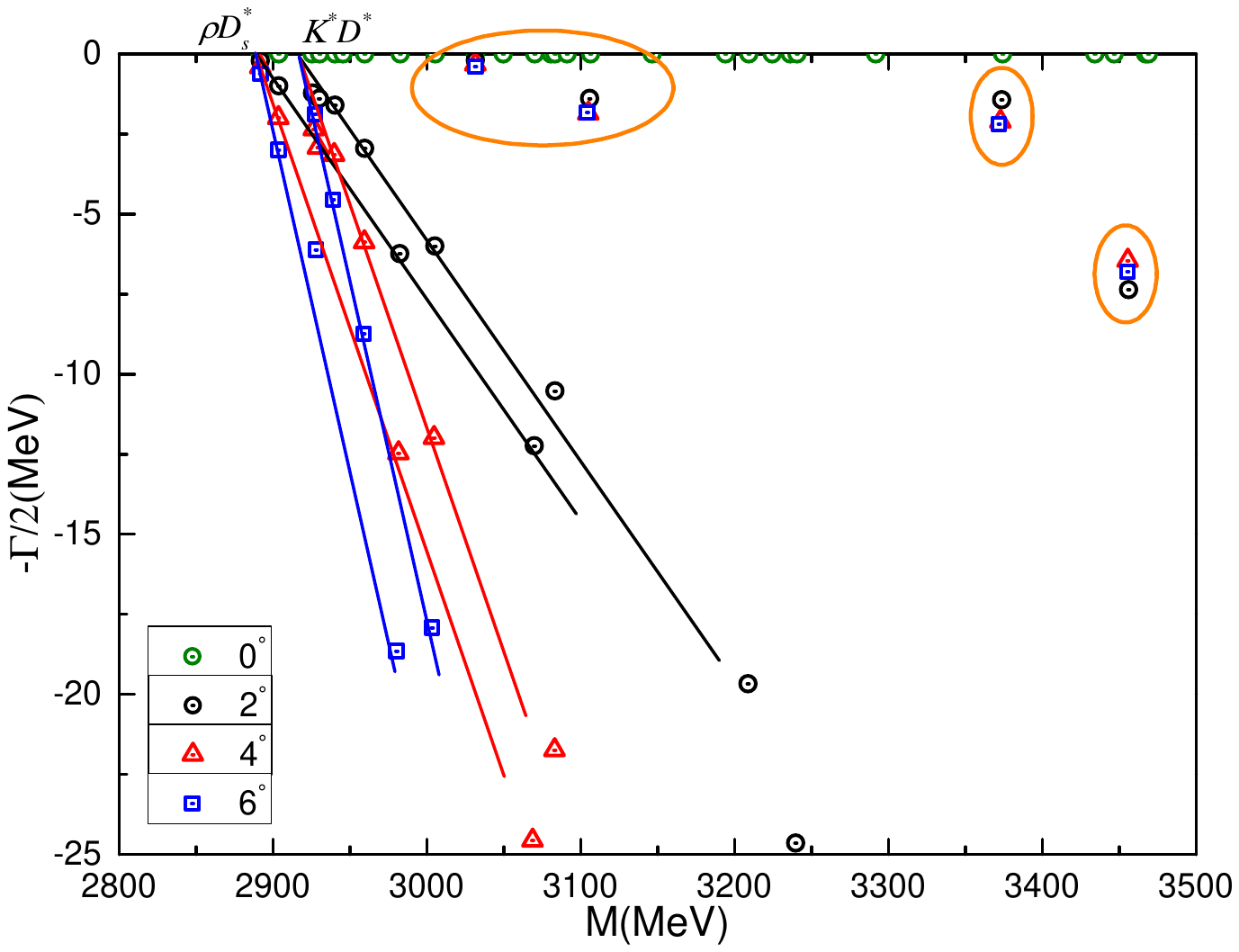}
\caption{\label{PP6} The complete coupled-channels calculation of $\bar{q}q\bar{s}c$ tetraquark system with $I(J^P)=1(2^+)$ quantum numbers.}
\end{figure}

\begin{table}[!t]
\caption{\label{GresultR6} Compositeness of exotic resonances obtained in a complete coupled-channel calculation in the $1(2^+)$ state of $\bar{q}q\bar{s}c$ tetraquark. Results are similarly organized as those in Table~\ref{GresultR1}.}
\begin{ruledtabular}
\begin{tabular}{lccc}
Resonance       & \multicolumn{3}{c}{Structure} \\[2ex]
$3031+i0.7$   & \multicolumn{3}{c}{$r_{q\bar{q}}:1.02$;\,\,\,\,\,$r_{\bar{q}\bar{q}}:1.07$;\,\,\,\,\,$r_{c\bar{q}}:0.94$;\,\,\,\,\,$r_{qc}:1.00$} \\
& \multicolumn{3}{c}{$S$: 16.4\%;\, $H$: 14.1\%;\, $Di$: 27.0\%;\, $K$: 42.5\%}\\[1.5ex]
$3105+i3.7$   & \multicolumn{3}{c}{$r_{q\bar{q}}:1.26$;\,\,\,\,\,$r_{\bar{q}\bar{q}}:1.33$;\,\,\,\,\,$r_{c\bar{q}}:0.99$;\,\,\,\,\,$r_{qc}:1.26$} \\
& \multicolumn{3}{c}{$S$: 4.0\%;\, $H$: 21.6\%;\, $Di$: 28.0\%;\, $K$: 46.4\%}\\[1.5ex]
$3373+i4.2$   & \multicolumn{3}{c}{$r_{q\bar{q}}:1.51$;\,\,\,\,\,$r_{\bar{q}\bar{q}}:1.61$;\,\,\,\,\,$r_{c\bar{q}}:1.45$;\,\,\,\,\,$r_{qc}:1.55$} \\
& \multicolumn{3}{c}{$S$: 8.0\%;\, $H$: 10.6\%;\, $Di$: 26.7\%;\, $K$: 54.7\%}\\[1.5ex]
$3455+i12.9$   & \multicolumn{3}{c}{$r_{q\bar{q}}:1.68$;\,\,\,\,\,$r_{\bar{q}\bar{q}}:1.54$;\,\,\,\,\,$r_{c\bar{q}}:1.30$;\,\,\,\,\,$r_{qc}:1.45$} \\
& \multicolumn{3}{c}{$S$: 6.6\%;\, $H$: 9.8\%;\, $Di$: 39.4\%;\, $K$: 44.2\%}
\end{tabular}
\end{ruledtabular}
\end{table}

{\bf The $\bm{I(J^P)=1(2^+)}$ sector:} Twelve channels are considered in the highest spin and isospin tetraquark state, and our results are listed in Table~\ref{GresultCC6}. Firstly, bound states are not found neither in single channel calculations nor in coupled-channel cases. The lowest channel is $\rho D^*_s$ with the theoretical threshold value of $2887$ MeV, and another dimeson structure $K^* D^*$ in the color-singlet channel is at $2924$ MeV. Masses of other channels with exotic configurations are generally in an energy region of $3.1-3.2$ GeV, and each of the lowest coupled mass in one specific structure is $\sim 3.16$ GeV.

The complete coupled-channels calculation using CSM is shown in Fig.~\ref{PP6}. Particularly, two scattering states of $\rho D^*_s$ and $K^* D^*$ are well presented within $2.8-3.5$ GeV. However, there are four stable poles above threshold lines. In Table~\ref{GresultR6} one can find the resonance masses, widths and wavefunction configurations. Moreover, one can guess the compact tetraquark structure, with size about $1.0-1.6$ fm, for the four resonances at $3031+i0.7$, $3105+i3.7$, $3373+i4.2$ and $3455+i12.9$ MeV, respectively. Besides, there are strong couplings among singlet-, hidden-color, diquark-antidiquark and K-type channels for these resonances. Both $\rho D^*_s$ and $K^* D^*$ are suggested to be golden decay channels.


\subsection{The $\mathbf{\bar{q}q\bar{s}b}$ tetraquarks}

Three spin-parity states, $J^P=0^+$, $1^+$ and $2^+$, with isospin $I=0$ and $1$, are investigated for the $\bar{q}q\bar{s}b$ tetraquark system. Several narrow resonances are obtained in each $I(J^P)$ quantum numbers. Details of the calculation as well as the related discussion can be found below.


\begin{table}[!t]
\caption{\label{GresultCC7} Lowest-lying $\bar{q}q\bar{s}b$ tetraquark states with $I(J^P)=0(0^+)$ calculated within the real range formulation of the chiral quark model. Results are similarly organized as those in Table~\ref{GresultCC1} (unit: MeV).}
\begin{ruledtabular}
\begin{tabular}{lcccc}
~~Channel   & Index & $\chi_J^{\sigma_i}$;~$\chi_I^{f_j}$;~$\chi_k^c$ & $M$ & Mixed~~ \\
        &   &$[i; ~j; ~k]$ &  \\[2ex]
$(\eta B_s)^1 (5915)$          & 1  & [1;~1;~1]  & $6044$ & \\
$(\omega B^*_s)^1 (6197)$  & 2  & [2;~1;~1]   & $6096$ &  \\
$(K B)^1 (5774)$          & 3  & [1;~1;~1]  & $5759$ & \\
$(K^* B^*)^1 (6217)$  & 4  & [2;~1;~1]   & $6226$ & $5759$ \\[2ex]
$(\eta B_s)^8$          & 5  & [1;~1;~2]  & $6613$ & \\
$(\omega B^*_s)^8$  & 6  & [2;~1;~2]   & $6440$ &  \\
$(K B)^8$          & 7  & [1;~1;~2]  & $6361$ & \\
$(K^* B^*)^8$  & 8  & [2;~1;~2]   & $6498$ & $6292$ \\[2ex]
$(qb)(\bar{q}\bar{s})$      & 9   & [3;~2;~4]  & $6501$ & \\
$(qb)(\bar{q}\bar{s})$      & 10   & [3;~3;~3]  & $6226$ & \\
$(qb)^*(\bar{q}\bar{s})^*$  & 11  & [4;~2;~3]   & $6486$ & \\
$(qb)^*(\bar{q}\bar{s})^*$  & 12  & [4;~3;~4]   & $6414$ & $6113$ \\[2ex]
$K_1$  & 13  & [5;~1;~5]   & $6403$ & \\
  & 14  & [6;~1;~5]   & $6585$ & \\
  & 15  & [5;~1;~6]   & $6358$ & \\
  & 16  & [6;~1;~6]   & $6387$ & $6281$ \\[2ex]
$K_2$  & 17  & [7;~1;~7]   & $6427$ & \\
  & 18  & [8;~1;~7]   & $6531$ & \\
  & 19  & [7;~1;~8]   & $6520$ & \\
  & 20  & [8;~1;~8]   & $6411$ & $6210$ \\[2ex]
$K_3$  & 21  & [9;~2;~10]   & $6485$ & \\
  & 22  & [9;~3;~9]   & $6382$ & \\
  & 23  & [10;~2;~9]   & $6478$ & \\
  & 24  & [10;~3;~10]   & $6221$ & $6038$ \\[2ex]
\multicolumn{4}{c}{Complete coupled-channels:} & $5759$
\end{tabular}
\end{ruledtabular}
\end{table}

\begin{figure}[!t]
\includegraphics[clip, trim={3.0cm 1.9cm 3.0cm 1.0cm}, width=0.45\textwidth]{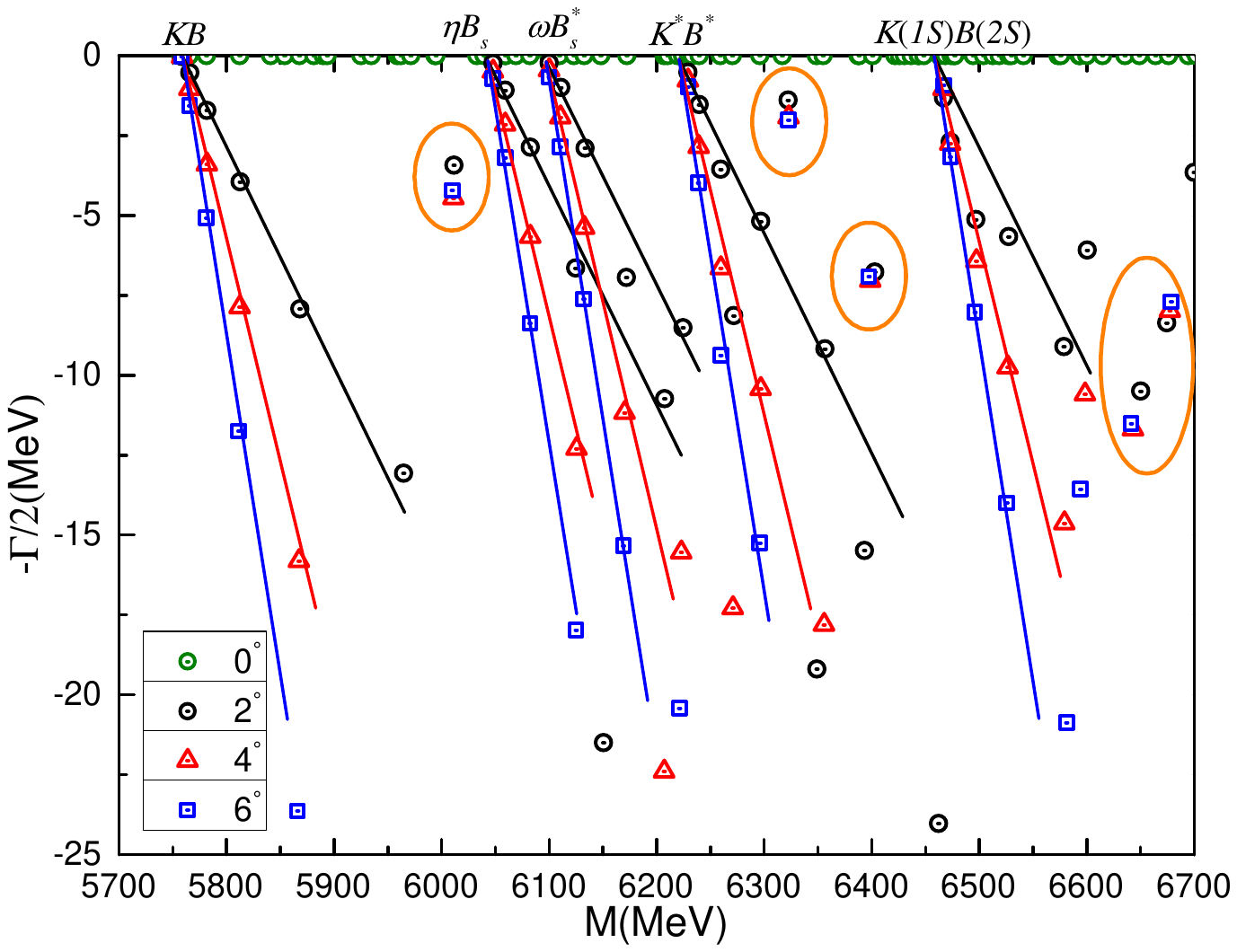}
\caption{\label{PP7} The complete coupled-channels calculation of $\bar{q}q\bar{s}b$ tetraquark system with $I(J^P)=0(0^+)$ quantum numbers.}
\end{figure}

\begin{table}[!t]
\caption{\label{GresultR7} Compositeness of exotic resonances obtained in a complete coupled-channel calculation in the $0(0^+)$ state of $\bar{q}q\bar{s}b$ tetraquark. Results are similarly organized as those in Table~\ref{GresultR1}.}
\begin{ruledtabular}
\begin{tabular}{lccc}
Resonance       & \multicolumn{3}{c}{Structure} \\[2ex]
$6011+i9.0$   & \multicolumn{3}{c}{$r_{q\bar{q}}:1.54$;\,\,\,\,\,$r_{\bar{q}\bar{q}}:1.53$;\,\,\,\,\,$r_{b\bar{q}}:0.66$;\,\,\,\,\,$r_{qb}:1.48$} \\
& \multicolumn{3}{c}{$S$: 19.5\%;\, $H$: 7.5\%;\, $Di$: 14.4\%;\, $K$: 58.6\%}\\[1.5ex]
$6323+i4.0$   & \multicolumn{3}{c}{$r_{q\bar{q}}:1.16$;\,\,\,\,\,$r_{\bar{q}\bar{q}}:1.24$;\,\,\,\,\,$r_{b\bar{q}}:0.94$;\,\,\,\,\,$r_{qb}:1.18$} \\
& \multicolumn{3}{c}{$S$: 16.9\%;\, $H$: 7.6\%;\, $Di$: 21.9\%;\, $K$: 53.6\%}\\[1.5ex]
$6397+i13.8$   & \multicolumn{3}{c}{$r_{q\bar{q}}:0.96$;\,\,\,\,\,$r_{\bar{q}\bar{q}}:1.81$;\,\,\,\,\,$r_{b\bar{q}}:1.74$;\,\,\,\,\,$r_{qb}:1.76$} \\
& \multicolumn{3}{c}{$S$: 7.8\%;\, $H$: 17.6\%;\, $Di$: 25.4\%;\, $K$: 49.2\%}\\[1.5ex]
$6643+i23.4$   & \multicolumn{3}{c}{$r_{q\bar{q}}:1.35$;\,\,\,\,\,$r_{\bar{q}\bar{q}}:1.61$;\,\,\,\,\,$r_{b\bar{q}}:1.31$;\,\,\,\,\,$r_{qb}:1.52$} \\
& \multicolumn{3}{c}{$S$: 13.1\%;\, $H$: 21.4\%;\, $Di$: 31.1\%;\, $K$: 34.4\%}\\[1.5ex]
$6678+i16.0$   & \multicolumn{3}{c}{$r_{q\bar{q}}:1.54$;\,\,\,\,\,$r_{\bar{q}\bar{q}}:1.60$;\,\,\,\,\,$r_{b\bar{q}}:1.33$;\,\,\,\,\,$r_{qb}:1.52$} \\
& \multicolumn{3}{c}{$S$: 17.9\%;\, $H$: 27.8\%;\, $Di$: 21.2\%;\, $K$: 33.1\%}
\end{tabular}
\end{ruledtabular}
\end{table}

{\bf The $\bm{I(J^P)=0(0^+)}$ sector:} In Table~\ref{GresultCC7}, one can see that 24 channels are investigated in this case. Firstly, concerning the four color-singlet channels, which include $\eta B_s$, $\omega B^*_s$, $K B$ and $K^* B^*$, the lowest mass is $5759$ MeV. This is just the theoretical threshold value of $K B$, and the other channels are also unbound. Furthermore, the lowest-lying channels of hidden-color, diquark-antidiquark and K-type configurations are generally located in a mass region $6.2-6.6$ GeV. When coupled-channel computations are performed in each kind of structure, the scattering nature of $K B$ channel in di-meson structure is still obtained. Besides, possible color resonances in $K_3$ and diquark-antidiquark structures are obtained at $6.04$ and $6.11$ GeV, respectively. The lowest masses of other three configurations are $\sim 6.2$ GeV. Finally, in a real-range computation, which is performed by including all of the above channels, the lowest mass of $\bar{q}q\bar{s}b$ tetraquark system remains at the $K B$ theoretical threshold, $5759$ MeV.

In the next step, the complete coupled-channels case is studied in a complex-range formulation. Figure~\ref{PP7} shows the distribution of complex energies within $5.7-6.7$ GeV. In particular, five scattering states, which are the ground state of $K B$, $\eta B_s$, $\omega B^*_s$ and $K^* B^*$, and the first radial excitation of $K(1S) B(2S)$, are well presented. Nevertheless, five resonances are also found and circled in Fig.~\ref{PP7}, their complex energies are $6011+i9$, $6323+i4$, $6397+i13.8$, $6643+i23.4$ and $6678+i16$ MeV, respectively.

Details about the properties of such resonances are listed in Table~\ref{GresultR7}. Firstly, there are strong couplings among the singlet-, hidden-color, diquark-antidiquark and K-type channels. Secondly, their sizes are less than $1.9$ fm. Particularly, the two lower resonances have sizes within $1.1-1.5$ fm, and the other three are extended about $1.3-1.8$ fm. Thirdly, for the lowest resonance at $6.0$ GeV, the golden decay channel is the $K B$; the $\omega B^*_s$ and $K^* B^*$ channels are dominant meson-meson components for the two resonances at $6.3$ GeV, while the remaining two resonances at $6.6$ GeV can be confirmed in $K B$ and $K^* B^*$ channels.


\begin{table}[!t]
\caption{\label{GresultCC8} Lowest-lying $\bar{q}q\bar{s}b$ tetraquark states with $I(J^P)=0(1^+)$ calculated within the real range formulation of the chiral quark model. Results are similarly organized as those in Table~\ref{GresultCC1} (unit: MeV).}
\begin{ruledtabular}
\begin{tabular}{lcccc}
~~Channel   & Index & $\chi_J^{\sigma_i}$;~$\chi_I^{f_j}$;~$\chi_k^c$ & $M$ & Mixed~~ \\
        &   &$[i; ~j; ~k]$ &  \\[2ex]
$(\eta B^*_s)^1 (5963)$          & 1  & [1;~1;~1]  & $6089$ & \\
$(\omega B_s)^1 (6149)$  & 2  & [2;~1;~1]   & $6051$ &  \\
$(\omega B^*_s)^1 (6197)$  & 3  & [3;~1;~1]   & $6096$ &  \\
$(K B^*)^1 (5819)$      & 4  & [1;~1;~1]   & $5800$ &  \\
$(K^* B)^1 (6172)$      & 5  & [2;~1;~1]  & $6185$ & \\
$(K^* B^*)^1 (6217)$  & 6  & [3;~1;~1]   & $6226$ & $5800$ \\[2ex]
$(\eta B^*_s)^8$  & 7  & [1;~1;~2]  & $6614$ & \\
$(\omega B_s)^8$    & 8  & [2;~1;~2]   & $6496$ &  \\
$(\omega B^*_s)^8$  & 9  & [3;~1;~2]   & $6470$ &  \\
$(K B^*)^8$      & 10  & [1;~1;~2]   & $6361$ &  \\
$(K^* B)^8$      & 11 & [2;~1;~2]  & $6360$ & \\
$(K^* B^*)^8$      & 12  & [3;~1;~2]   & $6497$ & $6315$ \\[2ex]
$(qb)(\bar{q}\bar{s})^*$      & 13   & [4;~2;~4]  & $6498$ & \\
$(qb)(\bar{q}\bar{s})^*$      & 14   & [4;~3;~3]  & $6239$ & \\
$(qb)^*(\bar{q}\bar{s})$      & 15   & [5;~2;~3]  & $6474$ & \\
$(qb)^*(\bar{q}\bar{s})$      & 16  & [5;~3;~4]   & $6473$ & \\
$(qb)^*(\bar{q}\bar{s})^*$  & 17  & [6;~2;~3]   & $6466$ &  \\
$(qb)^*(\bar{q}\bar{s})^*$  & 18  & [6;~3;~4]   & $6421$ & $6140$ \\[2ex]
$K_1$  & 19  & [7;~1;~5]   & $6480$ & \\
  & 20  & [8;~1;~5]   & $6410$ & \\
  & 21  & [9;~1;~5]   & $6588$ & \\
  & 22  & [7;~1;~6]   & $6351$ & \\
  & 23  & [8;~1;~6]   & $6352$ & \\
  & 24  & [9;~1;~6]   & $6405$ & $6299$ \\[2ex]
$K_2$  & 25  & [10;~1;~7]   & $6420$ & \\
  & 26  & [11;~1;~7]   & $6502$ & \\
  & 27  & [12;~1;~7]   & $6491$ & \\
  & 28  & [10;~1;~8]   & $6353$ & \\
  & 29  & [11;~1;~8]   & $6521$ & \\
  & 30  & [12;~1;~8]   & $6512$ & $6254$ \\[2ex]
$K_3$  & 31  & [13;~2;~10]   & $6389$ & \\
  & 32  & [13;~3;~9]   & $6386$ & \\
  & 33  & [14;~2;~10]   & $6488$ & \\
  & 34  & [14;~3;~9]   & $6398$ & \\
  & 35  & [15;~2;~10]   & $6474$ & \\
  & 36  & [15;~3;~9]   & $6231$ & $6063$ \\[2ex]
\multicolumn{4}{c}{Complete coupled-channels:} & $5800$
\end{tabular}
\end{ruledtabular}
\end{table}

\begin{figure}[!t]
\includegraphics[clip, trim={3.0cm 1.9cm 3.0cm 1.0cm}, width=0.45\textwidth]{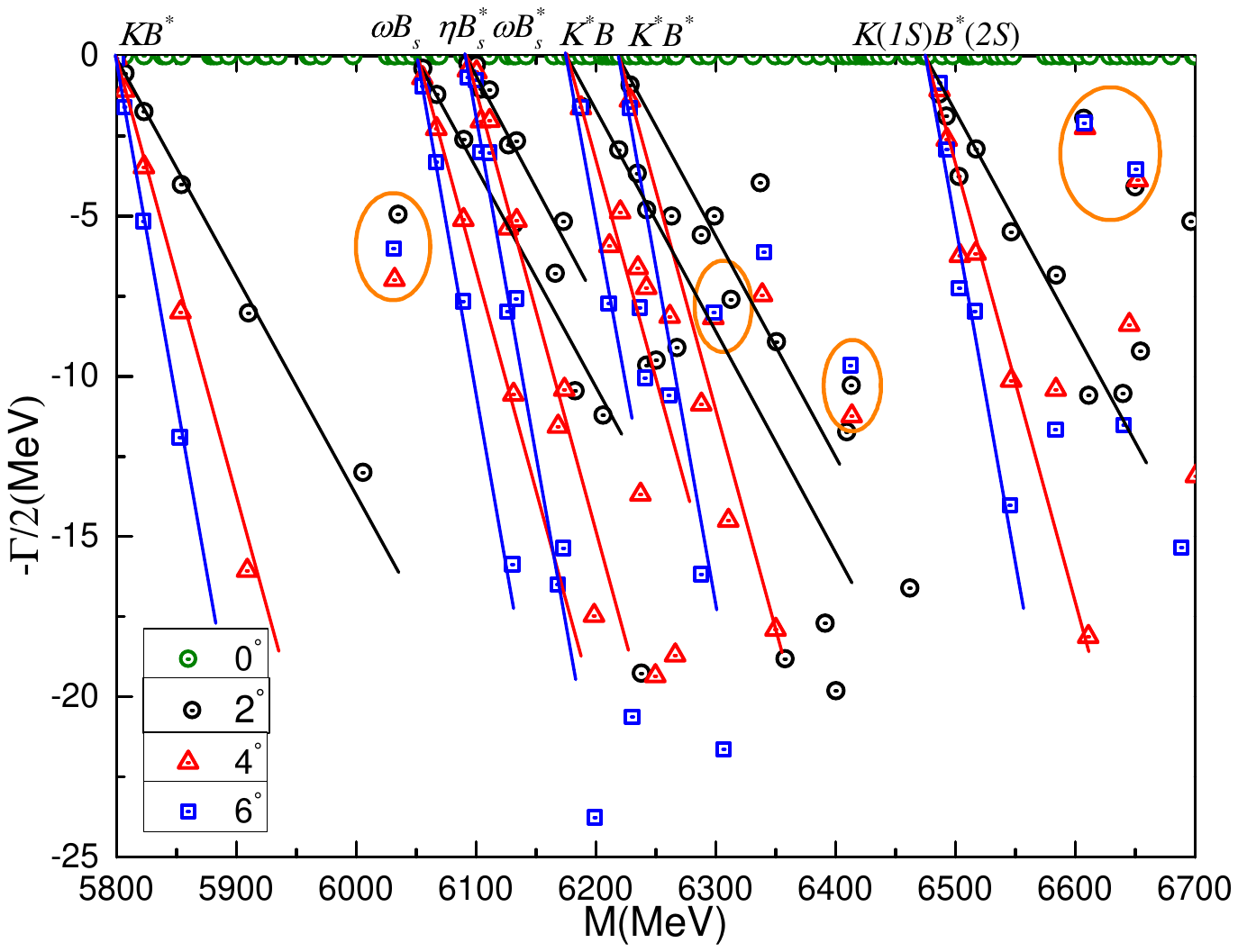}
\caption{\label{PP8} The complete coupled-channels calculation of $\bar{q}q\bar{s}b$ tetraquark system with $I(J^P)=0(1^+)$ quantum numbers.}
\end{figure}

\begin{table}[!t]
\caption{\label{GresultR8} Compositeness of exotic resonances obtained in a complete coupled-channel calculation in the $0(1^+)$ state of $\bar{q}q\bar{s}b$ tetraquark. Results are similarly organized as those in Table~\ref{GresultR1}.}
\begin{ruledtabular}
\begin{tabular}{lccc}
Resonance       & \multicolumn{3}{c}{Structure} \\[2ex]
$6031+i12.0$   & \multicolumn{3}{c}{$r_{q\bar{q}}:1.85$;\,\,\,\,\,$r_{\bar{q}\bar{q}}:1.87$;\,\,\,\,\,$r_{b\bar{q}}:0.73$;\,\,\,\,\,$r_{qb}:1.81$} \\
& \multicolumn{3}{c}{$S$: 21.3\%;\, $H$: 10.1\%;\, $Di$: 16.6\%;\, $K$: 52.0\%}\\[1.5ex]
$6298+i16.0$   & \multicolumn{3}{c}{$r_{q\bar{q}}:1.95$;\,\,\,\,\,$r_{\bar{q}\bar{q}}:1.94$;\,\,\,\,\,$r_{b\bar{q}}:0.91$;\,\,\,\,\,$r_{qb}:1.86$} \\
& \multicolumn{3}{c}{$S$: 27.4\%;\, $H$: 10.6\%;\, $Di$: 10.2\%;\, $K$: 51.8\%}\\[1.5ex]
$6413+i20.6$   & \multicolumn{3}{c}{$r_{q\bar{q}}:1.22$;\,\,\,\,\,$r_{\bar{q}\bar{q}}:1.55$;\,\,\,\,\,$r_{b\bar{q}}:0.80$;\,\,\,\,\,$r_{qb}:1.60$} \\
& \multicolumn{3}{c}{$S$: 16.6\%;\, $H$: 17.3\%;\, $Di$: 21.9\%;\, $K$: 44.2\%}\\[1.5ex]
$6607+i4.2$   & \multicolumn{3}{c}{$r_{q\bar{q}}:1.56$;\,\,\,\,\,$r_{\bar{q}\bar{q}}:1.84$;\,\,\,\,\,$r_{b\bar{q}}:1.70$;\,\,\,\,\,$r_{qb}:1.75$} \\
& \multicolumn{3}{c}{$S$: 17.9\%;\, $H$: 25.2\%;\, $Di$: 12.6\%;\, $K$: 44.3\%}\\[1.5ex]
$6652+i7.7$   & \multicolumn{3}{c}{$r_{q\bar{q}}:1.61$;\,\,\,\,\,$r_{\bar{q}\bar{q}}:1.93$;\,\,\,\,\,$r_{b\bar{q}}:1.67$;\,\,\,\,\,$r_{qb}:1.78$} \\
& \multicolumn{3}{c}{$S$: 14.8\%;\, $H$: 20.6\%;\, $Di$: 27.4\%;\, $K$: 37.2\%}
\end{tabular}
\end{ruledtabular}
\end{table}

{\bf The $\bm{I(J^P)=0(1^+)}$ sector:} Table~\ref{GresultCC8} summarizes our results in the real-range formalism. Firstly, the $\eta B^*_s$, $\omega B^{(*)}_s$ and $K^{(*)} B^{(*)}$ channels in both color-singlet and -octet cases are considered. The lowest mass is $5.8$ GeV, which is just the $K B^*$ theoretical threshold value. Moreover, no bound states in the other meson-meson channels are found, and the six hidden-color channels are generally located within the energy range $6.3-6.6$ GeV. This energy region is also shared by the diquark-antidiquark and three K-type channels. In partially coupled-channel calculations, the coupling effect is weak in singlet- and hidden-color channels, their lowest masses are $5.80$ and $6.32$ GeV, respectively. Therefore, the bound state is still unavailable. Although there are strong couplings in diquark-antidiquark and three K-type channels, and the mass shift for the lowest-lying channel is $50-170$ MeV, bound state is again not obtained. This result also holds for the fully coupled-channel calculation.

In a further complex analysis of the complete coupled-channel, five resonances are obtained and they are indicated in Fig.~\ref{PP8}. One can see, besides the seven scattering states of $K^{(*)} B^{(*)}$, $\omega B^{(*)}_s$ and $\eta B^*_s$ in the energy region $5.8-6.7$ GeV, the five stable poles circled, with complex energies given by $6031+i12$, $6298+i16$, $6413+i20.6$, $6607+i4.2$ and $6652+i7.7$ MeV, respectively.

Table~\ref{GresultR8} lists the calculated properties of the resonances in order to elucidate their nature. Firstly, strong coupling effects of different tetraquark configurations are reflected. Meanwhile, these resonances have sizes of about $1.6-1.9$ fm, except for the one at $6.4$ GeV whose size is less than $1.6$ fm. Finally, the lowest resonance at $6.03$ GeV is suggested to be experimentally studied in the $K B^*$ decay channel. The $K^* B$ and $K^* B^*$ are dominant two-body strong decay channels for the $6.29$ and $6.41$ GeV resonances whereas the $K B^*$ is the golden channel for the remaining two resonances at $6.6$ GeV.


\begin{table}[!t]
\caption{\label{GresultCC9} Lowest-lying $\bar{q}q\bar{s}b$ tetraquark states with $I(J^P)=0(2^+)$ calculated within the real range formulation of the chiral quark model. Results are similarly organized as those in Table~\ref{GresultCC1} (unit: MeV).}
\begin{ruledtabular}
\begin{tabular}{lcccc}
~~Channel   & Index & $\chi_J^{\sigma_i}$;~$\chi_I^{f_j}$;~$\chi_k^c$ & $M$ & Mixed~~ \\
        &   &$[i; ~j; ~k]$ &  \\[2ex]
$(\omega B^*_s)^1 (6197)$  & 1  & [1;~1;~1]   & $6096$ &  \\
$(K^* B^*)^1 (6217)$  & 2  & [1;~1;~1]   & $6226$ & $6096$ \\[2ex]
$(\omega B^*_s)^8$  & 3  & [1;~1;~2]   & $6524$ &  \\
$(K^* B^*)^8$  & 4  & [1;~1;~2]   & $6497$ & $6411$ \\[2ex]
$(qb)^*(\bar{q}\bar{s})^*$  & 5  & [1;~2;~3]   & $6422$ & \\
$(qb)^*(\bar{q}\bar{s})^*$  & 6  & [1;~3;~4]   & $6435$ & $6413$ \\[2ex]
$K_1$  & 7  & [1;~1;~5]   & $6489$ & \\
  & 8  & [1;~1;~6]   & $6363$ & $6362$ \\[2ex]
$K_2$  & 9  & [1;~1;~7]   & $6455$ & \\
  & 10  & [1;~1;~8]   & $6458$ & $6400$ \\[2ex]
$K_3$  & 11  & [1;~2;~10]   & $6403$ & \\
  & 12   & [1;~3;~9]   & $6397$ & $6387$ \\[2ex]
\multicolumn{4}{c}{Complete coupled-channels:} & $6096$
\end{tabular}
\end{ruledtabular}
\end{table}

\begin{table}[!t]
\caption{\label{GresultR9} Compositeness of exotic resonances obtained in a complete coupled-channel calculation in the $0(2^+)$ state of $\bar{q}q\bar{s}b$ tetraquark. Results are similarly organized as those in Table~\ref{GresultR1}.}
\begin{ruledtabular}
\begin{tabular}{lccc}
Resonance       & \multicolumn{3}{c}{Structure} \\[2ex]
$6239+i0.8$   & \multicolumn{3}{c}{$r_{q\bar{q}}:1.45$;\,\,\,\,\,$r_{\bar{q}\bar{q}}:1.50$;\,\,\,\,\,$r_{b\bar{q}}:0.93$;\,\,\,\,\,$r_{qb}:1.43$} \\
& \multicolumn{3}{c}{$S$: 19.1\%;\, $H$: 10.0\%;\, $Di$: 12.8\%;\, $K$: 58.1\%}\\[1.5ex]
$6314+i3.5$   & \multicolumn{3}{c}{$r_{q\bar{q}}:1.53$;\,\,\,\,\,$r_{\bar{q}\bar{q}}:1.52$;\,\,\,\,\,$r_{b\bar{q}}:0.80$;\,\,\,\,\,$r_{qb}:1.44$} \\
& \multicolumn{3}{c}{$S$: 22.4\%;\, $H$: 6.4\%;\, $Di$: 3.5\%;\, $K$: 67.7\%}\\[1.5ex]
$6619+i4.0$   & \multicolumn{3}{c}{$r_{q\bar{q}}:1.42$;\,\,\,\,\,$r_{\bar{q}\bar{q}}:1.44$;\,\,\,\,\,$r_{b\bar{q}}:1.27$;\,\,\,\,\,$r_{qb}:1.34$} \\
& \multicolumn{3}{c}{$S$: 17.3\%;\, $H$: 25.6\%;\, $Di$: 14.0\%;\, $K$: 43.1\%}\\[1.5ex]
$6664+i6.9$   & \multicolumn{3}{c}{$r_{q\bar{q}}:1.32$;\,\,\,\,\,$r_{\bar{q}\bar{q}}:1.53$;\,\,\,\,\,$r_{b\bar{q}}:1.34$;\,\,\,\,\,$r_{qb}:1.41$} \\
& \multicolumn{3}{c}{$S$: 12.1\%;\, $H$: 18.0\%;\, $Di$: 32.2\%;\, $K$: 37.7\%}
\end{tabular}
\end{ruledtabular}
\end{table}

\begin{figure}[!t]
\includegraphics[width=0.49\textwidth, trim={2.3cm 2.0cm 2.0cm 1.0cm}]{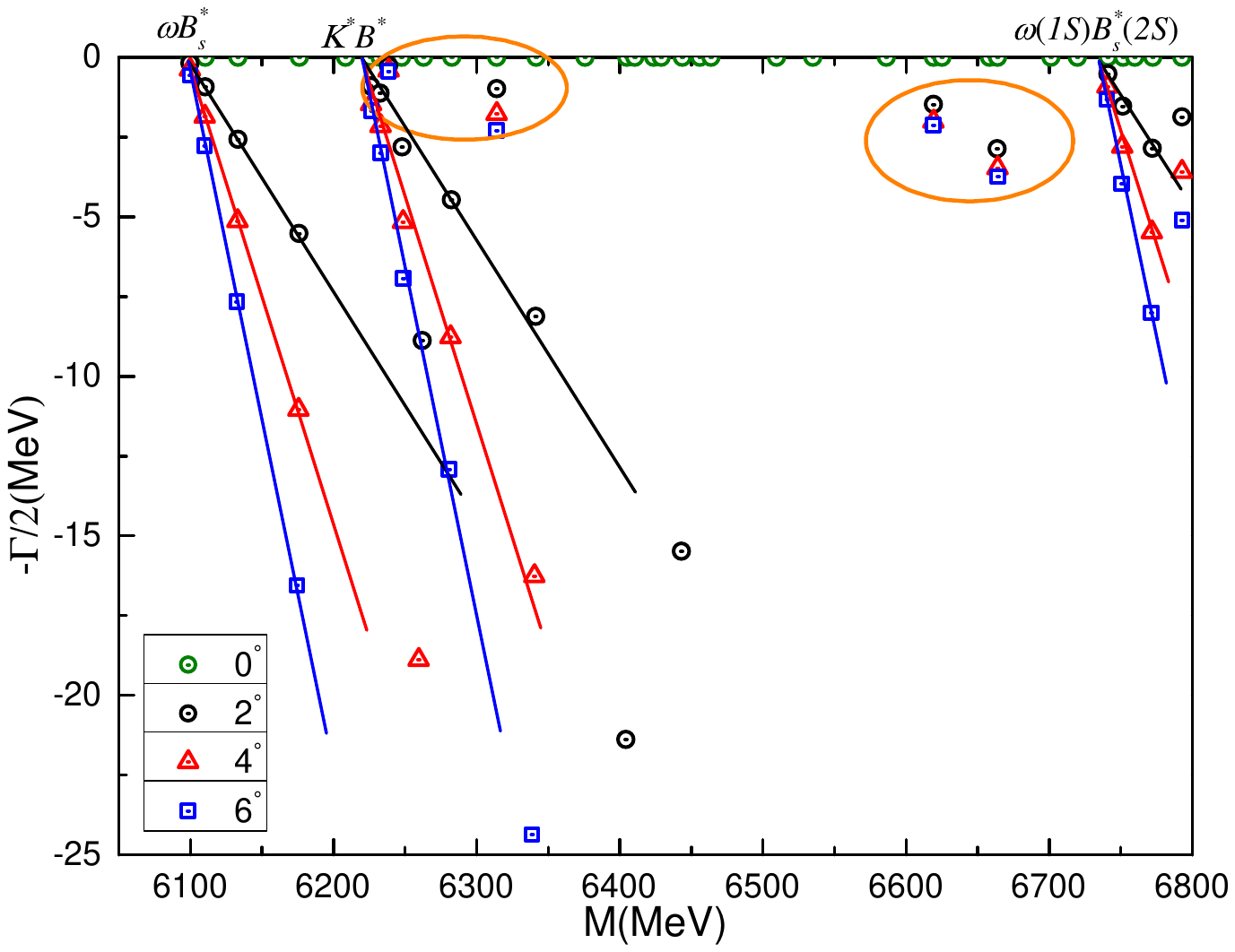}
\caption{\label{PP9} The complete coupled-channels calculation of $\bar{q}q\bar{s}b$ tetraquark system with $I(J^P)=0(2^+)$ quantum numbers.}
\end{figure}

{\bf The $\bm{I(J^P)=0(2^+)}$ sector:} The $\omega B^*_s$ and $K^* B^*$ in both color-singlet and hidden-color channels, along with two diquark-antidiquark, and six K-type channels are considered in this case. First of all, bound state is not found in single- and coupled-channel computations. The lowest-lying channel is the $\omega B^*_s$ scattering state. Additionally, other channels with exotic color structures are located in $6.36-6.52$ GeV. When a coupled-channels calculation is performed in each specific configuration, the lowest mass of all of them is $\sim 6.4$ GeV.

Furthermore, Fig.~\ref{PP9} presents results in the fully coupled-channel case using the CSM. Therein, scattering states of $\omega B^*_s$, $K^* B^*$ and $\omega(1S) B^*_s(2S)$ are well shown; additionally, four narrow resonances are also found. In Table~\ref{GresultR9} one can find their resonance parameters: $6239+i0.8$, $6314+i3.5$, $6619+i4$ and $6664+i6.9$ MeV, respectively. Moreover, the size of these four resonances is around $1.4$ fm; the proportions of singlet-, hidden-color, diquark-antidiquark and K-type channels are comparable. They can be further confirmed experimentally in $\omega B^*_s$ and $K^* B^*$ channels.


\begin{figure}[!t]
\includegraphics[width=0.49\textwidth, trim={2.3cm 2.0cm 2.0cm 1.0cm}]{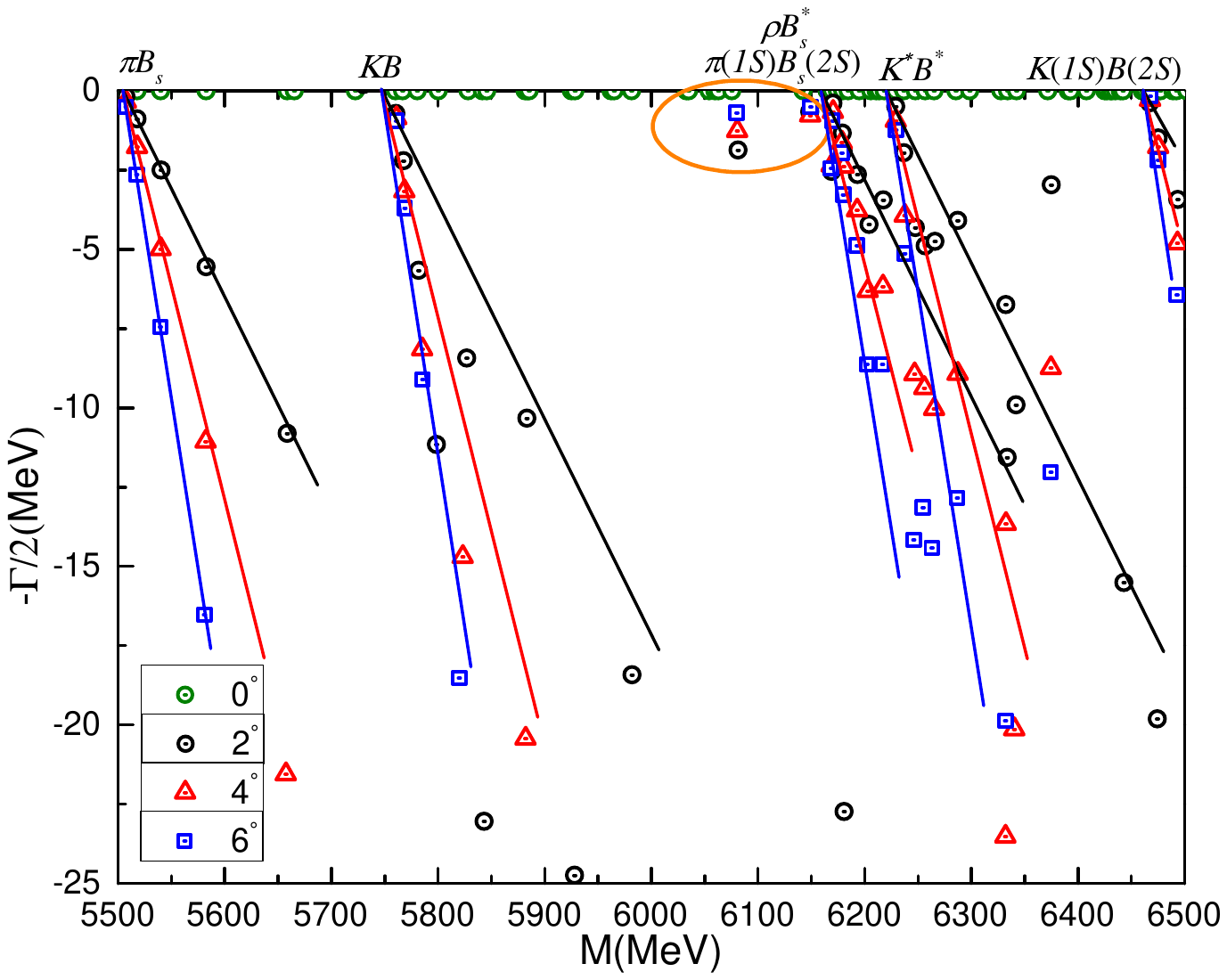}
\caption{\label{PP10} The complete coupled-channels calculation of $\bar{q}q\bar{s}b$ tetraquark system with $I(J^P)=1(0^+)$ quantum numbers.}
\end{figure}

\begin{table}[!t]
\caption{\label{GresultCC10} Lowest-lying $\bar{q}q\bar{s}b$ tetraquark states with $I(J^P)=1(0^+)$ calculated within the real range formulation of the chiral quark model. Results are similarly organized as those in Table~\ref{GresultCC1} (unit: MeV).}
\begin{ruledtabular}
\begin{tabular}{lcccc}
~~Channel   & Index & $\chi_J^{\sigma_i}$;~$\chi_I^{f_j}$;~$\chi_k^c$ & $M$ & Mixed~~ \\
        &   &$[i; ~j; ~k]$ &  \\[2ex]
$(\pi B_s)^1 (5507)$          & 1  & [1;~1;~1]  & $5504$ & \\
$(\rho B^*_s)^1 (6185)$  & 2  & [2;~1;~1]   & $6172$ &  \\
$(K B)^1 (5774)$          & 3  & [1;~1;~1]  & $5759$ & \\
$(K^* B^*)^1 (6217)$  & 4  & [2;~1;~1]   & $6226$ & $5504$ \\[2ex]
$(\pi B_s)^8$          & 5  & [1;~1;~2]  & $6494$ & \\
$(\rho B^*_s)^8$  & 6  & [2;~1;~2]   & $6499$ &  \\
$(K B)^8$          & 7  & [1;~1;~2]  & $6361$ & \\
$(K^* B^*)^8$  & 8  & [2;~1;~2]   & $6420$ & $6243$ \\[2ex]
$(qb)(\bar{q}\bar{s})$      & 9   & [3;~2;~4]  & $6501$ & \\
$(qb)(\bar{q}\bar{s})$      & 10   & [3;~3;~3]  & $6226$ & \\
$(qb)^*(\bar{q}\bar{s})^*$  & 11  & [4;~2;~3]   & $6298$ & \\
$(qb)^*(\bar{q}\bar{s})^*$  & 12  & [4;~3;~4]   & $6302$ & $6123$ \\[2ex]
$K_1$  & 13  & [5;~1;~5]   & $6468$ & \\
  & 14  & [6;~1;~5]   & $6463$ & \\
  & 15  & [5;~1;~6]   & $6449$ & \\
  & 16  & [6;~1;~6]   & $5792$ & $5780$ \\[2ex]
$K_2$  & 17  & [7;~1;~7]   & $6315$ & \\
  & 18  & [8;~1;~7]   & $6531$ & \\
  & 19  & [7;~1;~8]   & $6417$ & \\
  & 20  & [8;~1;~8]   & $6411$ & $6200$ \\[2ex]
$K_3$  & 21  & [9;~2;~10]   & $6355$ & \\
  & 22  & [9;~3;~9]   & $6203$ & \\
  & 23  & [10;~2;~9]   & $6478$ & \\
  & 24  & [10;~3;~10]   & $6221$ & $6029$ \\[2ex]
\multicolumn{4}{c}{Complete coupled-channels:} & $5504$
\end{tabular}
\end{ruledtabular}
\end{table}

\begin{table}[!t]
\caption{\label{GresultR10} Compositeness of exotic resonances obtained in a complete coupled-channel calculation in the $1(0^+)$ state of $\bar{q}q\bar{s}b$ tetraquark. Results are similarly organized as those in Table~\ref{GresultR1}.}
\begin{ruledtabular}
\begin{tabular}{lccc}
Resonance       & \multicolumn{3}{c}{Structure} \\[2ex]
$6080+i2.5$   & \multicolumn{3}{c}{$r_{q\bar{q}}:0.95$;\,\,\,\,\,$r_{\bar{q}\bar{q}}:1.37$;\,\,\,\,\,$r_{b\bar{q}}:1.24$;\,\,\,\,\,$r_{qb}:1.27$} \\
& \multicolumn{3}{c}{$S$: 18.3\%;\, $H$: 10.6\%;\, $Di$: 12.8\%;\, $K$: 58.3\%}\\[1.5ex]
$6149+i1.6$   & \multicolumn{3}{c}{$r_{q\bar{q}}:1.09$;\,\,\,\,\,$r_{\bar{q}\bar{q}}:1.13$;\,\,\,\,\,$r_{b\bar{q}}:1.10$;\,\,\,\,\,$r_{qb}:1.00$} \\
& \multicolumn{3}{c}{$S$: 18.5\%;\, $H$: 10.5\%;\, $Di$: 8.5\%;\, $K$: 62.5\%}
\end{tabular}
\end{ruledtabular}
\end{table}

{\bf The $\bm{I(J^P)=1(0^+)}$ sector:} Firstly, all channels listed in Table~\ref{GresultCC10} are investigated in the real-range computations, and no bound state is obtained. The lowest-lying scattering state is $\pi B_s$ with a theoretical threshold value $5504$ MeV. The other three di-meson scattering states are $\rho B^*_s$, $K B$ and $K^* B^*$. Furthermore, masses of these four meson-meson structures in hidden-color channels are around $6.4$ GeV, and it is also similar for the $K_1$ channels, except for one at $5.79$ GeV. As for the diquark-antidiquark, $K_2$ and $K_3$ channels, they are generally located in $6.2-6.5$ GeV. In coupled-channel studies, which include six partial and one complete channels calculations, strong and weak coupling effects are both presented. In particular, channel couplings in the hidden-color, diquark-antidiquark, $K_2$ and $K_3$ configurations are strong, and they present $100-170$ MeV mass shifts. Their lowest masses are $6.24$, $6.12$, $6.20$ and $6.03$ GeV, respectively. However, the coupling is weak in color-singlet, $K_1$ and fully-coupled channels calculation. Accordingly, the scattering nature of $\pi B_s$ remains unchanged.

In a further step, the complex-range study is carried on the complete coupled-channels case. Six scattering states are plotted in Fig.~\ref{PP10}, and they are the ground states of $\pi B_s$, $K B$, $\rho B^*_s$ and $K^* B^*$, and the first radial excitations of $\pi(1S) B_s(2S)$ and $K(1S) B(2S)$. Moreover, within an energy region $5.5-6.5$ GeV, two narrow resonance poles are obtained.

The compositeness of these resonances is listed in Table~\ref{GresultR10}. Firstly, the complex energies of the two resonances read as $6080+i2.5$ and $6149+i1.6$ MeV, respectively. Meanwhile, they are compact $\bar{q}q\bar{s}b$ tetraquark structures, whose sizes are $\sim 1.2$ fm. Couplings among color-singlet, -octet, diquark-antidiquark and K-type channels are strong. Both $\pi B_s$ and $K B$ are their golden channels to be discover. 


\begin{figure}[!t]
\includegraphics[clip, trim={3.0cm 1.9cm 3.0cm 1.0cm}, width=0.45\textwidth]{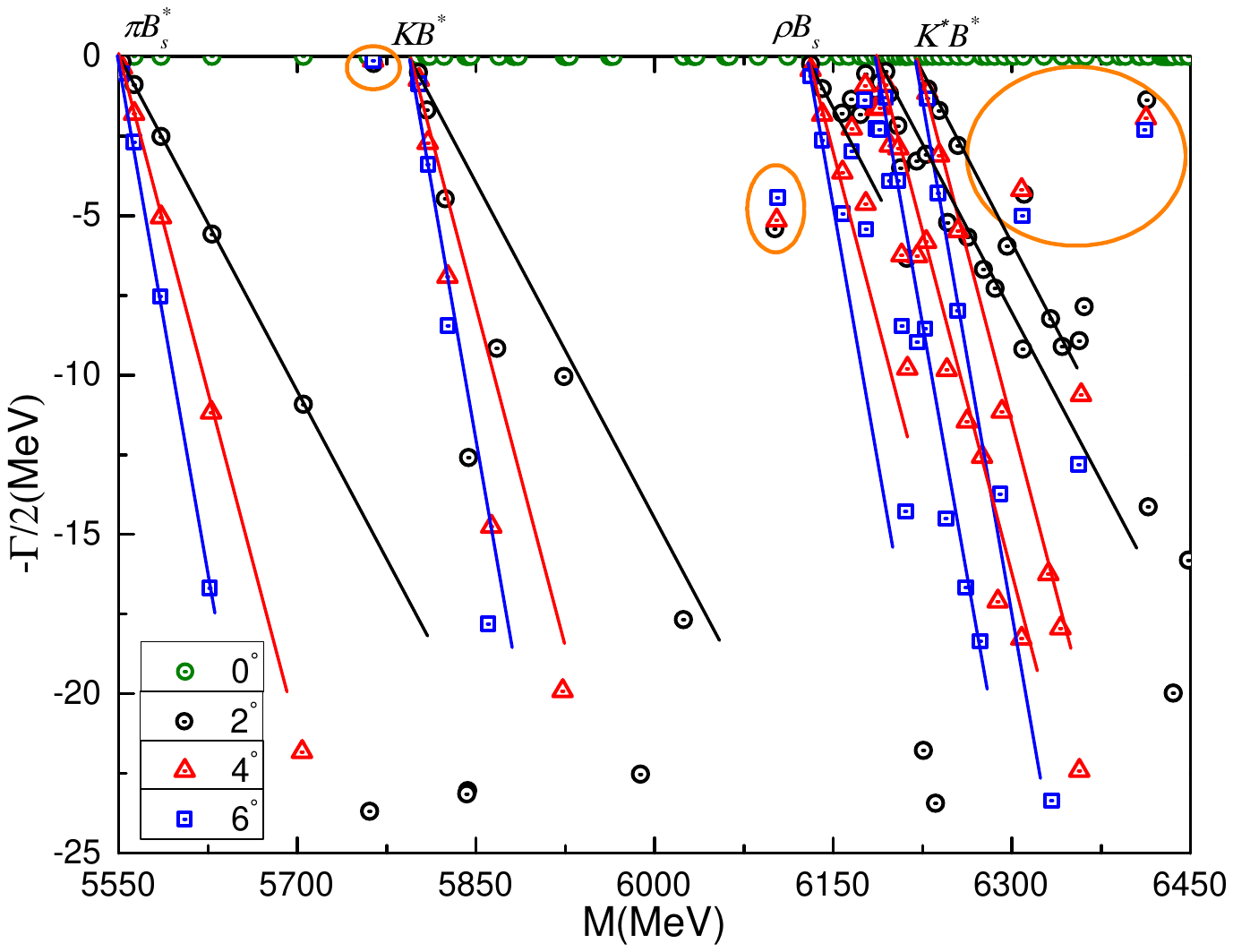} \\[1ex]
\includegraphics[clip, trim={3.0cm 1.9cm 3.0cm 1.0cm}, width=0.45\textwidth]{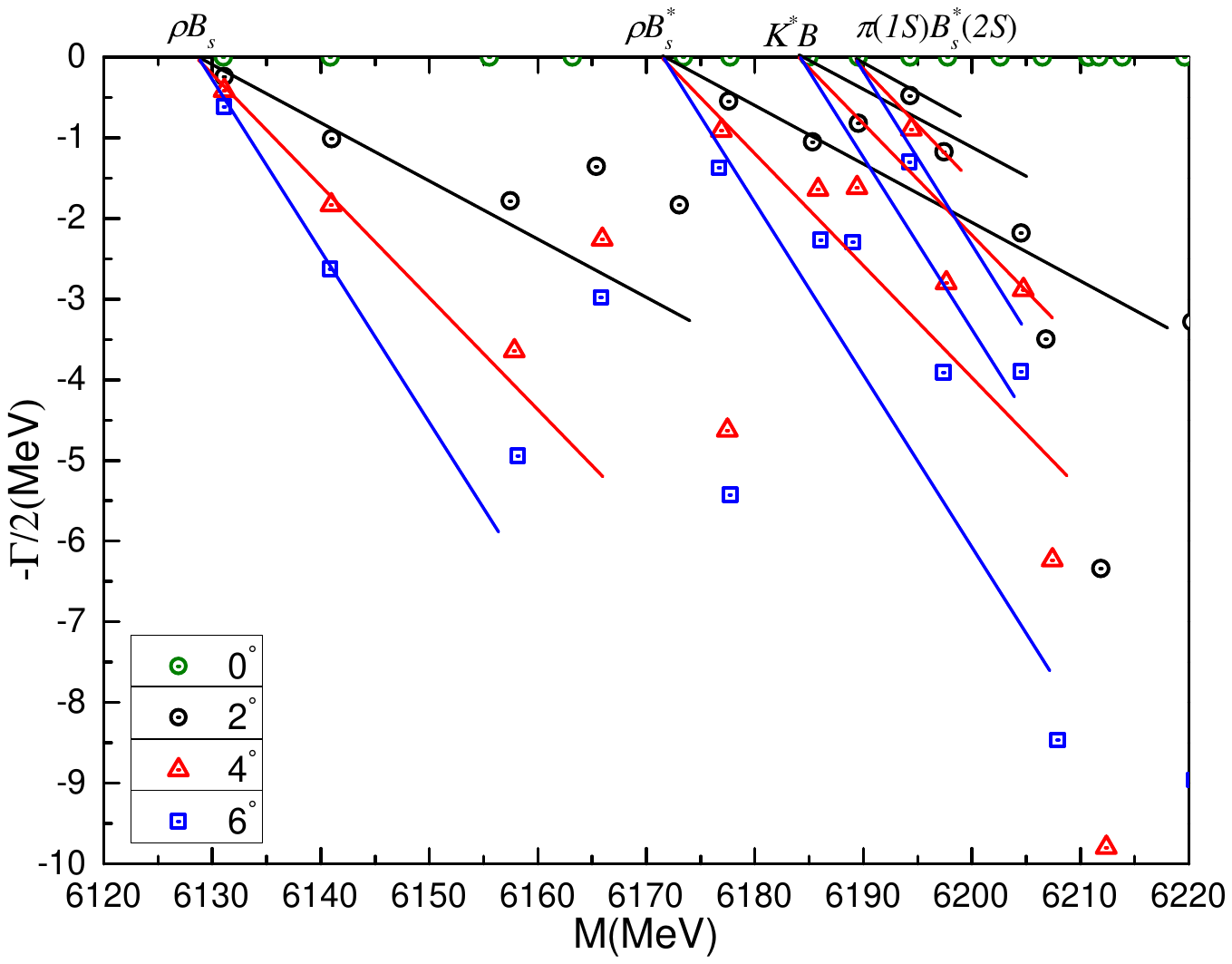}
\caption{\label{PP11} {\it Top panel:} The complete coupled-channels calculation of $\bar{q}q\bar{s}b$ tetraquark system with $I(J^P)=1(1^+)$ quantum numbers. Particularly, the bottom panel is enlarged parts of dense energy region from $6.12\,\text{GeV}$ to $6.22\,\text{GeV}$.}
\end{figure}

\begin{table}[!t]
\caption{\label{GresultCC11} Lowest-lying $\bar{q}q\bar{s}b$ tetraquark states with $I(J^P)=1(1^+)$ calculated within the real range formulation of the chiral quark model. Results are similarly organized as those in Table~\ref{GresultCC1} (unit: MeV).}
\begin{ruledtabular}
\begin{tabular}{lcccc}
~~Channel   & Index & $\chi_J^{\sigma_i}$;~$\chi_I^{f_j}$;~$\chi_k^c$ & $M$ & Mixed~~ \\
        &   &$[i; ~j; ~k]$ &  \\[2ex]
$(\pi B^*_s)^1 (5555)$          & 1  & [1;~1;~1]  & $5549$ & \\
$(\rho B_s)^1 (6137)$  & 2  & [2;~1;~1]   & $6127$ &  \\
$(\rho B^*_s)^1 (6185)$  & 3  & [3;~1;~1]   & $6172$ &  \\
$(K B^*)^1 (5819)$      & 4  & [1;~1;~1]   & $5800$ &  \\
$(K^* B)^1 (6172)$      & 5  & [2;~1;~1]  & $6185$ & \\
$(K^* B^*)^1 (6217)$  & 6  & [3;~1;~1]   & $6226$ & $5549$ \\[2ex]
$(\pi B^*_s)^8$  & 7  & [1;~1;~2]  & $6496$ & \\
$(\rho B_s)^8$    & 8  & [2;~1;~2]   & $6550$ &  \\
$(\rho B^*_s)^8$  & 9  & [3;~1;~2]   & $6526$ &  \\
$(K B^*)^8$      & 10  & [1;~1;~2]   & $6361$ &  \\
$(K^* B)^8$      & 11 & [2;~1;~2]  & $6360$ & \\
$(K^* B^*)^8$      & 12  & [3;~1;~2]   & $6460$ & $6275$ \\[2ex]
$(qb)(\bar{q}\bar{s})^*$      & 13   & [4;~2;~4]  & $6498$ & \\
$(qb)(\bar{q}\bar{s})^*$      & 14   & [4;~3;~3]  & $6239$ & \\
$(qb)^*(\bar{q}\bar{s})$      & 15   & [5;~2;~3]  & $6474$ & \\
$(qb)^*(\bar{q}\bar{s})$      & 16  & [5;~3;~4]   & $6473$ & \\
$(qb)^*(\bar{q}\bar{s})^*$  & 17  & [6;~2;~3]   & $6415$ &  \\
$(qb)^*(\bar{q}\bar{s})^*$  & 18  & [6;~3;~4]   & $6370$ & $6159$ \\[2ex]
$K_1$  & 19  & [7;~1;~5]   & $6536$ & \\
  & 20  & [8;~1;~5]   & $6474$ & \\
  & 21  & [9;~1;~5]   & $6466$ & \\
  & 22  & [7;~1;~6]   & $6442$ & \\
  & 23  & [8;~1;~6]   & $6443$ & \\
  & 24  & [9;~1;~6]   & $5809$ & $5800$ \\[2ex]
$K_2$  & 25  & [10;~1;~7]   & $6468$ & \\
  & 26  & [11;~1;~7]   & $6405$ & \\
  & 27  & [12;~1;~7]   & $6491$ & \\
  & 28  & [10;~1;~8]   & $6415$ & \\
  & 29  & [11;~1;~8]   & $6418$ & \\
  & 30  & [12;~1;~8]   & $6512$ & $6261$ \\[2ex]
$K_3$  & 31  & [13;~2;~10]   & $6454$ & \\
  & 32  & [13;~3;~9]   & $6451$ & \\
  & 33  & [14;~2;~10]   & $6358$ & \\
  & 34  & [14;~3;~9]   & $6224$ & \\
  & 35  & [15;~2;~10]   & $6474$ & \\
  & 36  & [15;~3;~9]   & $6231$ & $6065$ \\[2ex]
\multicolumn{4}{c}{Complete coupled-channels:} & $5549$
\end{tabular}
\end{ruledtabular}
\end{table}

\begin{table}[!t]
\caption{\label{GresultR11} Compositeness of exotic resonances obtained in a complete coupled-channel calculation in the $1(1^+)$ state of $\bar{q}q\bar{s}b$ tetraquark. Results are similarly organized as those in Table~\ref{GresultR1}.}
\begin{ruledtabular}
\begin{tabular}{lccc}
Resonance       & \multicolumn{3}{c}{Structure} \\[2ex]
$5764+i0.4$   & \multicolumn{3}{c}{$r_{q\bar{q}}:0.89$;\,\,\,\,\,$r_{\bar{q}\bar{q}}:0.89$;\,\,\,\,\,$r_{b\bar{q}}:0.78$;\,\,\,\,\,$r_{qb}:0.69$} \\
& \multicolumn{3}{c}{$S$: 29.6\%;\, $H$: 2.4\%;\, $Di$: 9.2\%;\, $K$: 58.8\%}\\[1.5ex]
$6103+i10.3$   & \multicolumn{3}{c}{$r_{q\bar{q}}:0.93$;\,\,\,\,\,$r_{\bar{q}\bar{q}}:1.73$;\,\,\,\,\,$r_{b\bar{q}}:1.58$;\,\,\,\,\,$r_{qb}:1.66$} \\
& \multicolumn{3}{c}{$S$: 17.6\%;\, $H$: 13.1\%;\, $Di$: 23.2\%;\, $K$: 46.1\%}\\[1.5ex]
$6308+i8.4$   & \multicolumn{3}{c}{$r_{q\bar{q}}:1.17$;\,\,\,\,\,$r_{\bar{q}\bar{q}}:1.65$;\,\,\,\,\,$r_{b\bar{q}}:1.40$;\,\,\,\,\,$r_{qb}:1.37$} \\
& \multicolumn{3}{c}{$S$: 11.4\%;\, $H$: 6.1\%;\, $Di$: 11.7\%;\, $K$: 70.8\%}\\[1.5ex]
$6413+i3.8$   & \multicolumn{3}{c}{$r_{q\bar{q}}:1.50$;\,\,\,\,\,$r_{\bar{q}\bar{q}}:1.63$;\,\,\,\,\,$r_{b\bar{q}}:1.55$;\,\,\,\,\,$r_{qb}:1.40$} \\
& \multicolumn{3}{c}{$S$: 15.6\%;\, $H$: 19.8\%;\, $Di$: 27.3\%;\, $K$: 37.3\%}
\end{tabular}
\end{ruledtabular}
\end{table}

{\bf The $\bm{I(J^P)=1(1^+)}$ sector:} Table~\ref{GresultCC11} lists 36 channels under consideration for this quantum state. In the meson-meson color-singlet channels, $\pi B^*_s$, $\rho B^{(*)}_s$, and $K^{(*)} B^{(*)}$ are calculated. The lowest channel is the scattering state of $\pi B^*_s$, and its mass is just the theoretical threshold value of $5549$ MeV. Besides, the scattering nature of $\pi B^*_s$ channel remains in partially and fully coupled-channel calculations. Other channels are also unbound. As for channels in the other five structures, which are hidden-color, diquark-antidiquark and K-types configurations, they are generally located in an energy region $6.2-6.5$ GeV, except for a $K_1$ channel at $5.8$ GeV. Additionally, when coupled-channels calculations are considered in each of these five structures, a weak coupling effect is obtained in $K_1$ channels, and the lowest coupled mass is still $5.8$ GeV. In contrast, there are strong coupling effects in other configurations. Nevertheless, they are still unstable excited states within $6.06-6.28$ GeV.

In the complete coupled-channels computation using CSM, the distribution of complex energies is plotted in Fig.~\ref{PP11}. Particularly, within an energy region $5.55-6.45$ GeV of the top panel of Fig.~\ref{PP11}, scattering states of $\pi B^*_s$, $K B^*$, $\rho B_s$ and $K^* B^*$ are well presented. Furthermore, there are dense distributions of energy dots at around $6.2$ GeV, hence an enlarged part from $6.12$ to $6.22$ GeV is plotted in the bottom panel. Therein, four scattering states of $\rho B^{(*)}_s$, $K^* B$ and $\pi(1S) B^*_s(2S)$ are shown too.

Apart from the obtained continuum states, four resonances are also found in the complex plane. Table~\ref{GresultR11} summarizes their calculated results. Firstly, the four stable poles read $5764+i0.4$, $6103+i10.3$, $6308+i8.4$ and $6413+i3.8$ MeV, respectively. Besides, color-singlet, diquark-antidiquark and K-type channels couplings are strong for the resonances. Compact structure, with size around $0.8$ fm, is obtained for the lowest resonance at $5.76$ GeV, and the dominant meson-meson component is $\pi B^*_s$. However, the other three resonances are loose structures with size $\sim 1.6$ fm. The golden channels of the second resonance at $6.1$ GeV are $\pi B^*_s$ and $K B^*$, while $\pi B^*_s$, $K^* B$ and $K^* B^*$ channels are suggested to be the dominant dimeson components for the other two higher resonances at $6.3$ and $6.4$ GeV, respectively.


\begin{figure}[!t]
\includegraphics[width=0.49\textwidth, trim={2.3cm 2.0cm 2.0cm 1.0cm}]{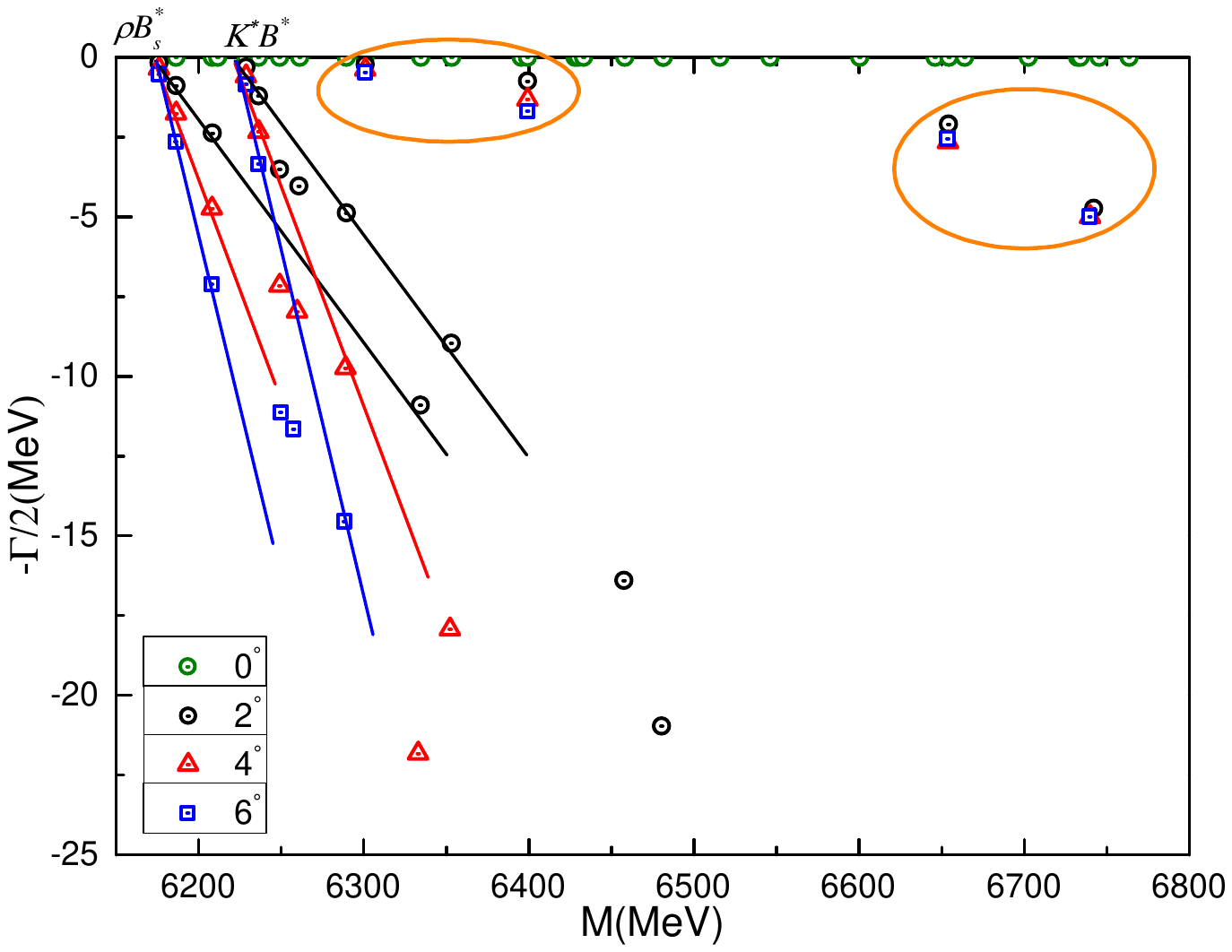}
\caption{\label{PP12} The complete coupled-channels calculation of $\bar{q}q\bar{s}b$ tetraquark system with $I(J^P)=1(2^+)$ quantum numbers.}
\end{figure}

\begin{table}[!t]
\caption{\label{GresultCC12} Lowest-lying $\bar{q}q\bar{s}b$ tetraquark states with $I(J^P)=1(2^+)$ calculated within the real range formulation of the chiral quark model. Results are similarly organized as those in Table~\ref{GresultCC1} (unit: MeV).}
\begin{ruledtabular}
\begin{tabular}{lcccc}
~~Channel   & Index & $\chi_J^{\sigma_i}$;~$\chi_I^{f_j}$;~$\chi_k^c$ & $M$ & Mixed~~ \\
        &   &$[i; ~j; ~k]$ &  \\[2ex]
$(\rho B^*_s)^1 (6185)$  & 1  & [1;~1;~1]   & $6172$ &  \\
$(K^* B^*)^1 (6217)$  & 2  & [1;~1;~1]   & $6226$ & $6172$ \\[2ex]
$(\rho B^*_s)^8$  & 3  & [1;~1;~2]   & $6575$ &  \\
$(K^* B^*)^8$  & 4  & [1;~1;~2]   & $6531$ & $6448$ \\[2ex]
$(qb)^*(\bar{q}\bar{s})^*$  & 5  & [1;~2;~3]   & $6472$ & \\
$(qb)^*(\bar{q}\bar{s})^*$  & 6  & [1;~3;~4]   & $6479$ & $6462$ \\[2ex]
$K_1$  & 7  & [1;~1;~5]   & $6544$ & \\
  & 8  & [1;~1;~6]   & $6453$ & $6452$ \\[2ex]
$K_2$  & 9  & [1;~1;~7]   & $6500$ & \\
  & 10  & [1;~1;~8]   & $6510$ & $6451$ \\[2ex]
$K_3$  & 11  & [1;~2;~10]   & $6466$ & \\
  & 12   & [1;~3;~9]   & $6462$ & $6452$ \\[2ex]
\multicolumn{4}{c}{Complete coupled-channels:} & $6172$
\end{tabular}
\end{ruledtabular}
\end{table}

\begin{table}[!t]
\caption{\label{GresultR12} Compositeness of exotic resonances obtained in a complete coupled-channel calculation in the $1(2^+)$ state of $\bar{q}q\bar{s}b$ tetraquark. Results are similarly organized as those in Table~\ref{GresultR1}.}
\begin{ruledtabular}
\begin{tabular}{lccc}
Resonance       & \multicolumn{3}{c}{Structure} \\[2ex]
$6301+i0.8$   & \multicolumn{3}{c}{$r_{q\bar{q}}:1.03$;\,\,\,\,\,$r_{\bar{q}\bar{q}}:1.08$;\,\,\,\,\,$r_{b\bar{q}}:0.95$;\,\,\,\,\,$r_{qb}:1.00$} \\
& \multicolumn{3}{c}{$S$: 15.1\%;\, $H$: 11.5\%;\, $Di$: 13.9\%;\, $K$: 59.5\%}\\[1.5ex]
$6399+i2.6$   & \multicolumn{3}{c}{$r_{q\bar{q}}:1.16$;\,\,\,\,\,$r_{\bar{q}\bar{q}}:1.14$;\,\,\,\,\,$r_{b\bar{q}}:0.79$;\,\,\,\,\,$r_{qb}:1.02$} \\
& \multicolumn{3}{c}{$S$: 15.2\%;\, $H$: 15.9\%;\, $Di$: 11.8\%;\, $K$: 57.1\%}\\[1.5ex]
$6654+i5.3$   & \multicolumn{3}{c}{$r_{q\bar{q}}:1.48$;\,\,\,\,\,$r_{\bar{q}\bar{q}}:1.50$;\,\,\,\,\,$r_{b\bar{q}}:1.34$;\,\,\,\,\,$r_{qb}:1.43$} \\
& \multicolumn{3}{c}{$S$: 10.4\%;\, $H$: 15.4\%;\, $Di$: 27.8\%;\, $K$: 46.4\%}\\[1.5ex]
$6740+i10.0$   & \multicolumn{3}{c}{$r_{q\bar{q}}:1.48$;\,\,\,\,\,$r_{\bar{q}\bar{q}}:1.42$;\,\,\,\,\,$r_{b\bar{q}}:1.16$;\,\,\,\,\,$r_{qb}:1.32$} \\
& \multicolumn{3}{c}{$S$: 4.9\%;\, $H$: 7.3\%;\, $Di$: 41.6\%;\, $K$: 46.2\%}
\end{tabular}
\end{ruledtabular}
\end{table}

{\bf The $\bm{I(J^P)=1(2^+)}$ sector:} Twelve channels listed in Table~\ref{GresultCC12} are studied for the highest spin and isospin state of $\bar{q}q\bar{s}b$ tetraquark. First of all, bound states are not found in three kinds of calculations: single channel, partially and fully coupled-channels. The lowest scattering state is the $\rho B^*_s$, with a theoretical threshold value of $6172$ MeV; another one is the $6226$ MeV threshold value of $K^* B^*$ channel. Furthermore, masses of channels in five exotic color structures are generally located in the energy region $6.4-6.5$ GeV. The lowest mass obtained within a coupled-channels calculation for each specific configuration is always located at $\sim 6.45$ GeV.

Figure~\ref{PP12} shows the distribution of complex energies in the complete coupled-channels computation with the CSM employed. Within $6.15-6.80$ GeV energy region, the $\rho B^*_s$ and $K^* B^*$ scattering states are clearly presented. Meanwhile, four stable resonance poles are also obtained, and they are circled in the complex plane.

Table~\ref{GresultR12} lists properties of these resonances. In particular, their complex energies are $6301+i0.8$, $6399+i2.6$, $6654+i5.3$ and $6740+i10$ MeV, respectively. Couplings among color-singlet, -octet, diquark-antidiquark and K-type channels are strong for the first three resonances. However, there is only a strong coupling between diquark-antidiquark and K-type channels for the highest resonance. Moreover, compact $\bar{q}q\bar{s}b$ tetraquark structure is obtained for the four resonances because their sizes are less than $1.5$ fm. Finally, both $\rho B^*_s$ and $K^* B^*$ are the dominant meson-meson components of these exotic states.


\begin{table*}[!t]
\caption{\label{GresultCCT} Summary of resonance structures found in the $\bar{q}q\bar{s}Q$ $(q=u,\,d;\,Q=c,\,b)$ tetraquark systems. The first column shows the isospin, total spin and parity of each singularity. The second column refers to the dominant configuration components, particularly, $H$: hidden color, $Di$: diquark-antidiquark, $K$: K-type. Theoretical resonances are presented with the following notation: $E=M+i\Gamma$ in the last column (unit: MeV).}
\begin{ruledtabular}
\begin{tabular}{lcc}
\multicolumn{3}{c}{The $\bar{q}q\bar{s}c$ tetraquarks} \\[1ex]
~ $I(J^P)$ & Dominant Component   & Theoretical resonance~~ \\
\hline
~~$0(0^+)$  & $\omega D^*_s (10\%)+K^* D^*(11\%)+Di(13\%)+K(57\%)$   & $3006+i6.3$~~  \\[2ex]
~~$0(1^+)$    & $H(19\%)+Di(30\%)+K(43\%)$   & $3119+i19.8$~~ \\
               & $Di(35\%)+K(42\%)$   & $3292+i13.1$~~ \\
               & $Di(35\%)+K(50\%)$   & $3346+i22.0$~~ \\[2ex]
~~$0(2^+)$    & $\omega D^*_s (12\%)+K^* D^*(7\%)+Di(28\%)+K(45\%)$   & $2965+i0.5$~~ \\
                       & $\omega D^*_s (8\%)+K^* D^*(7\%)+H(13\%)+K(65\%)$   & $3026+i3.8$~~ \\
                       & $H(20\%)+Di(19\%)+K(48\%)$   & $3344+i3.3$~~ \\[2ex]
~~$1(0^+)$     & $H(19\%)+Di(30\%)+K(35\%)$   & $2770+i1.5$~~\\[2ex]
~~$1(1^+)$     & $\pi D^*_s (24\%)+Di(14\%)+K(56\%)$   & $2470+i1.0$~~\\
                       & $K^* D^* (10\%)+H(18\%)+Di(17\%)+K(49\%)$   & $3134+i2.0$~~\\[2ex]
~~$1(2^+)$    & $Di(27\%)+K(43\%)$   & $3031+i0.7$~~ \\
                      & $H(22\%)+Di(28\%)+K(46\%)$   & $3105+i3.7$~~ \\
                      & $Di(27\%)+K(55\%)$   & $3373+i4.2$~~ \\
                      & $Di(39\%)+K(44\%)$   & $3455+i12.9$~~ \\
\hline
\hline
\multicolumn{3}{c}{The $\bar{q}q\bar{s}b$ tetraquarks} \\[1ex]
~ $I(J^P)$ & Dominant Component   & Theoretical resonance~~ \\
\hline
~~$0(0^+)$    & $K B (17\%)+Di(14\%)+K(59\%)$ & $6011+i9.0$~~  \\
                      & $K^* B^* (10\%)+Di(22\%)+K(54\%)$   & $6323+i4.0$~~ \\
                      & $H(18\%)+Di(25\%)+K(49\%)$   & $6397+i13.8$~~ \\
                      & $H(21\%)+Di(31\%)+K(34\%)$   & $6643+i23.4$~~ \\
                      & $H(28\%)+Di(21\%)+K(33\%)$   & $6678+i16.0$~~\\[2ex]
~~$0(1^+)$    & $K B^* (17\%)+Di(17\%)+K(52\%)$   & $6031+i12.0$~~ \\
                      & $K^* B (10\%)+K^* B^* (12\%)+K(52\%)$   & $6298+i16.0$~~ \\
                      & $Di(22\%)+K(44\%)$   & $6413+i20.6$~~ \\
                      & $K B^* (11\%)+H(25\%)+K(44\%)$   & $6607+i4.2$~~ \\
                      & $H(21\%)+Di(27\%)+K(37\%)$   & $6652+i7.7$~~\\[2ex]
~~$0(2^+)$    & $\omega B^*_s (9\%)+K^* B^*(10\%)+Di(13\%)+K(58\%)$   & $6239+i0.8$~~ \\
                       & $\omega B^*_s (15\%)+K^* B^*(7\%)+K(68\%)$   & $6314+i3.5$~~ \\
                       & $\omega B^*_s (11\%)+K^* B^*(7\%)+H(26\%)+K(43\%)$   & $6619+i4.0$~~ \\
                       & $H(18\%)+Di(32\%)+K(38\%)$   & $6664+i6.9$~~ \\[2ex]
~~$1(0^+)$     & $\pi B_s (7\%)+K B(8\%)+Di(13\%)+K(58\%)$   & $6080+i2.5$~~\\
                       & $\pi B_s (8\%)+K B(5\%)+H(11\%)+K(63\%)$   & $6149+i1.6$~~\\[2ex]
~~$1(1^+)$    & $\pi B^*_s (27\%)+K(59\%)$   & $5764+i0.4$~~ \\
                      & $Di(23\%)+K(46\%)$   & $6103+i10.3$~~ \\
                      & $\pi B^*_s (4\%)+K^* B^* (5\%)+K(71\%)$   & $6308+i8.4$~~ \\
                      & $\pi B^*_s (6\%)+K^* B^*(4\%)+Di(27\%)+K(37\%)$   & $6413+i3.8$~~ \\[2ex]
~~$1(2^+)$    & $\rho B^*_s (6\%)+K^* B^* (9\%)+Di(14\%)+K(60\%)$   & $6301+i0.8$~~ \\
                      & $\rho B^*_s (13\%)+H(16\%)+K(57\%)$   & $6399+i2.6$~~ \\
                      & $Di(28\%)+K(46\%)$   & $6654+i5.3$~~ \\
                      & $Di(42\%)+K(46\%)$   & $6740+i10.0$~~
\end{tabular}
\end{ruledtabular}
\end{table*}

\section{Summary}
\label{sec:summary}

The $S$-wave $\bar{q}q\bar{s}Q$ $(q=u,\,d;\,Q=c,\,b)$ tetraquarks with spin-parity $J^P=0^+$, $1^+$ and $2^+$, and isospin $I=0$ and $1$, have been systematically investigated in a chiral quark model formalism. Furthermore, the color-singlet, -octet meson-meson configurations, diquark-antidiquark arrangements with their allowed color triplet-antitriplet and sextet-antisextet channels, and K-type configurations have been all considered. The four-body bound and resonant states have been determined by means of a highly efficient numerical approach: the Gaussian expansion method (GEM) supplemented with a complex-scaling analysis (CSM). Three kinds of computations have been generally presented: single channel, partially-coupled and fully-coupled channels.

Table~\ref{GresultCCT} summarizes our theoretical findings for the $\bar{q}q\bar{s}c$ and $\bar{q}q\bar{s}b$ tetraquark systems. The first column shows the quantum numbers $I(J^P)$, the second one expresses the dominant configurations and the third one lists the complex eigenenergies. Several conclusions are drawn below.

Firstly, the experimentally reported $T_{c\bar{s}}(2900)$ state can be well identified in the $I(J^P)=1(0^+)$ channel of the $\bar{q}q\bar{s}c$ tetraquark. It is a loosely-bound molecular structure with comparable components of $\pi D_s$ and $K D$; besides, there are also space in its wave function to allocate significant exotic color structures such as hidden-color, diquark-antidiquark and K-type channels. Furthermore, two extremely narrow resonances below $3.0$ GeV are also obtained in $0(2^+)$ and $1(1^+)$ channels. Their complex energies are $2470+i1.0$ and $2965+i0.5$ MeV, respectively. Meanwhile, several narrow resonances of the $\bar{q}q\bar{s}c$ system with different quantum numbers are also obtained within an energy region $3.0-3.4$ GeV. Generally, the coupling between hidden-color, diquark-antidiquark and K-type configurations is strong in all these cases.

In the $\bar{q}q\bar{s}b$ tetraquark systems, narrow resonances with masses located in the energy region $6.0-6.7$ GeV are obtained in all allowed $I(J^P)$ cases. The lowest resonance of the $\bar{q}q\bar{s}b$ system is found at $5.76$ GeV with quantum numbers $I(J^P)=1(1^+)$; note, however, that the predicted width is very small. Besides, it seems that strong couplings among exotic color structures are found. 

All of the identified exotic states are expected to be confirmed in future high-energy particle and nuclear experiments.


\begin{acknowledgments}
Work partially financed by National Natural Science Foundation of China under Grant Nos. 12305093, 11535005 and 11775118; Zhejiang Provincial Natural Science Foundation under Grant No. LQ22A050004; Ministerio Espa\~nol de Ciencia e Innovaci\'on under grant Nos. PID2019-107844GB-C22 and PID2022-140440NB-C22; the Junta de Andaluc\'ia under contract Nos. Operativo FEDER Andaluc\'ia 2014-2020 UHU-1264517, P18-FR-5057 and also PAIDI FQM-370.
\end{acknowledgments}


\bibliography{qqsQ}

\end{document}